\documentclass[apj]{emulateapj}

    \usepackage{amsmath,amsthm, amssymb, latexsym}
    	\usepackage{multirow}
    \usepackage{xspace}
    \usepackage{ifpdf}
    \usepackage{booktabs}
    \usepackage{rotating}
    \usepackage{longtable} 
    \usepackage{epstopdf} 
    \usepackage{textcomp}
    \usepackage[english]{babel}
    \usepackage[latin1]{inputenc}
	\newcommand{\bnabla}{\mbox{\boldmath $\nabla$}}
    \usepackage{hyperref}

 \hypersetup{
    pdftitle={Three-dimensional simulations of Magnetohydrodynamic waves in magnetized solar atmosphere},
    pdfauthor={G. Vigeesh}, 
    colorlinks=true,       
    linkcolor=black,          
    citecolor=blue,        
}

\shorttitle{3D simulations of MHD waves}
\shortauthors{Vigeesh et al.}

\begin{document}
\title{Three-dimensional simulations of Magnetohydrodynamic waves in magnetized solar atmosphere}

\author{{Vigeesh, G.}\altaffilmark{1,2};
Fedun, V.\altaffilmark{3};
Hasan, S. S.\altaffilmark{2};
Erd\'{e}lyi, R.\altaffilmark{3}}

\altaffiltext{1}{Department of Astronomy, New Mexico State University, 
Las Cruces,
NM, USA}
\altaffiltext{2}{Indian Institute of Astrophysics, Bangalore, India}
\altaffiltext{3}{SP$^{\text 2}$RC, Department of Applied Mathematics, 
University of
Sheffield, UK}

\begin{abstract}
We present results of three-dimensional numerical simulations of magnetohydrodynamic wave propagation in a solar magnetic flux tube. Our study aims at understanding the properties of a range of MHD wave modes generated by different photospheric motions. We consider two scenarios observed in the lower solar photosphere, namely, granular buffeting and vortex-like motion, among the simplest mechanism for the generation of waves within a strong, localized magnetic flux concentration. We show that, granular buffeting is likely to generate stronger slow and fast magneto-acoustic waves as compared to swirly motions. Correspondingly, the energy flux transported differ as a result of the driving motions. We also demonstrate that the waves generated by granular buffeting are likely to manifest in stronger emission in the chromospheric network. We argue that different mechanisms of wave generation are active during the evolution of a magnetic element in the intergranular lane, resulting in temporally varying emission at chromospheric heights.   
\end{abstract}


\section{Introduction}\label{s:introduction}

The highly dynamic solar photosphere is a rich source of waves, which are believed to be one of the plausible candidates responsible for heating the upper solar atmosphere
\citep[see a recent review by][in this context]{2009SSRv..149..229T}.  
Magnetic fields and magnetohydrodynamic (MHD) waves play an important role in the solar atmosphere's energy budget. With recent high-resolution observations using both spaceborne (Hinode/SOT, SoHO, SUNRISE) and ground-based (SST, DST/ROSA) telescopes and simulations by various groups 
\citep[e.g.][]{2009A&A...507L...9W,2011A&A...526A...5S}
showing that the solar photosphere harbours vortex motions at the downdraft regions of convective cells, here we aim to investigate the efficiency of these motions in the generation of a range of MHD modes and their transport of energy. Using three-dimensional simulations, we show that the excited MHD modes and the wave energy transported by these modes vary with different excitation scenarios.

Several authors have carried out studies of wave propagation in magnetic flux tubes. Some of the recent work on magnetoacoustic wave propagation by e.g. 
\cite{2008ApJ...680.1542H}, 
\cite{2008SoPh..251..589K},  
\cite{2009A&A...508..951V},  
and
\cite{2011SoPh..tmp..349V}  
considered magnetic network fields with a range of typical properties. They have highlighted the importance of mode coupling in the transport of energy by magneto-acoustic waves in a magnetised atmosphere. These studies were carried out in 2D, and they do provide a reliable picture of the actual processes taking place in the real atmosphere. A more recent work by
\cite{2011ApJ...727...17F} 
considers an open flux tube in the two-dimensional solar atmosphere with combination of realistic temperature profiles corresponding to different height ranges. These investigations bring out the importance of the transition region in influencing the propagation of magneto-acoustic waves.  
While a lot of information can still be unveiled in 2D with realistic atmosphere, similar to work carried out by
\cite{2011ApJ...727...17F}, 
a 3D approach to the problem is more appealing in terms of physical processes, although more expensive in terms of computational resources.
One of the major new features by allowing an extra dimension to the problem is the possibility of the excitation of the ``intermediate'' Alfv\'{e}n wave. The propagation of MHD waves in a three-dimensional atmosphere by the driving motions at the lower boundary has been studied by 
\cite{2007SoPh..246...41M}, 
\cite{2007A&A...467.1299E}, 
\cite{2009SoPh..258..219F}. 
They considered a uniform background magnetic field embedded in a realistic hydrodynamic model of a quiet Sun and have shown that magnetic field channels the photospheric disturbances all the way into the solar corona. These studies did not consider Alfv\'{e}n waves due to the non-torsional nature of the driving motion. Further studies of MHD simulations in three-dimensional models
\citep{2011AnGeo..29.1029F}
have, however, shown that the vortex motions in the solar photosphere
\citep{2011AnGeo..29..883S}
do generate different types of MHD modes, including torsional Alfv\'{e}n waves. In a recent paper, 
\cite{2011ApJ...740L..46F}, 
have demonstrated that torsional Alfv\'{e}n waves can reveal the magnetic structuring of chromospheric features, since the magnetic flux tubes act as a spatial frequency filter
\citep{2009Sci...323.1582J,2010ApJ...714.1637V}. 
These studies 
\citep[e.g.][]{2011SoPh..tmp..349V,2011ApJ...740L..46F}
highlight that the various MHD modes are a valuable tool for the magneto-seismic studies of the lower solar atmosphere. A considerable leap forward connecting the swirly lower solar atmosphere to the corona is made by 
\cite{2012Nature_Wedemeyer} 
where the role of Alfv\'en waves in channelling photospheric energy to the corona is demonstrated both observationally and confirmed by modelling. 

High-resolution observations using the Swedish Solar Telescope (SST) 
\citep{2008ApJ...687L.131B}
have revealed convectively driven vortex flows in the photosphere of the Sun. These flows result from downdrafts that occur in the regions where the flows converge at the boundary of convective cells. 
\cite{2008ApJ...687L.131B} 
found that the associated magnetic concentrations, visible as magnetic bright points (MBP) follow a spiral pattern as they are carried along by the vortical flow. 
\cite{2009A&A...507L...9W} 
have detected similar features in the core of the chromospheric line Ca II (854.2~nm) as a likely result of the disturbance caused in the magnetic fields due to the twisting of the underlying photospheric magnetic field.
New results from the SUNRISE experiment indicate that the vortex flows are a very common phenomena in the solar atmosphere
\citep{2010ApJ...723L.139B}. 
They are formed in localized convective downdrafts as a result of  angular momentum conservation. These observations also reveal vortex tubes
\citep{2010ApJ...723L.180S} 
with vorticity directed horizontal and are considered to cause asymmetric brightenings in the granular boundaries.
The direct detection of torsional Alfv\'{e}n oscillations associated with MBPs, as reported by 
\cite{2009Sci...323.1582J},
can be considered a likely consequence of the swirling of the magnetic-flux tubes as a result of the vortical motions.

Vortex-like features are also seen in large-eddy simulations of
\cite{2011A&A...526A...5S} 
pointing to the fact that these vortical motions are common feature in the magnetised photosphere as a result of a convective driver. This numerical prediction is now observationally confirmed by
\cite{2012Nature_Wedemeyer}.
Using realistic three-dimensional radiative hydrodynamic simulations, 
\cite{2011ApJ...727L..50K} 
show that interactions of vortices in the intergranular lanes excite acoustic waves into the overlying layers. 
\cite{2011A&A...533A.126M} 
have also reported the occurrence of vortices in surface convection simulations.

In this paper, we model individual flux tubes located in inter-granular lanes in an atmosphere representing the empirical temperature profile of the solar chromosphere. Convective motions are thought to excite waves in this medium, modelled here by means of a driver located in the photosphere.
Two different driving mechanisms are considered, viz. transverse uni-directional and torsional motions. The driving generates various MHD modes within the flux tube and acoustic waves in the ambient medium.
We study the propagation and other properties of MHD modes generated by these drivers.

Section~\ref{s:initial_model} discusses the construction of the initial magneto-hydrostatic equilibrium model and the properties of the model.
The excitation mechanisms are explained in Sec.~\ref{s:wave_excitation}. The numerical methods and boundary conditions are described in Sec.~\ref{s:numerical_methods}. In Sec.~\ref{s:results} we discuss the results and the conclusions are presented in Sec.~\ref{s:conclusion}.

\section{Initial Magneto-hydrostatic Model}\label{s:initial_model}

\begin{figure}
\includegraphics[width=0.95\columnwidth]{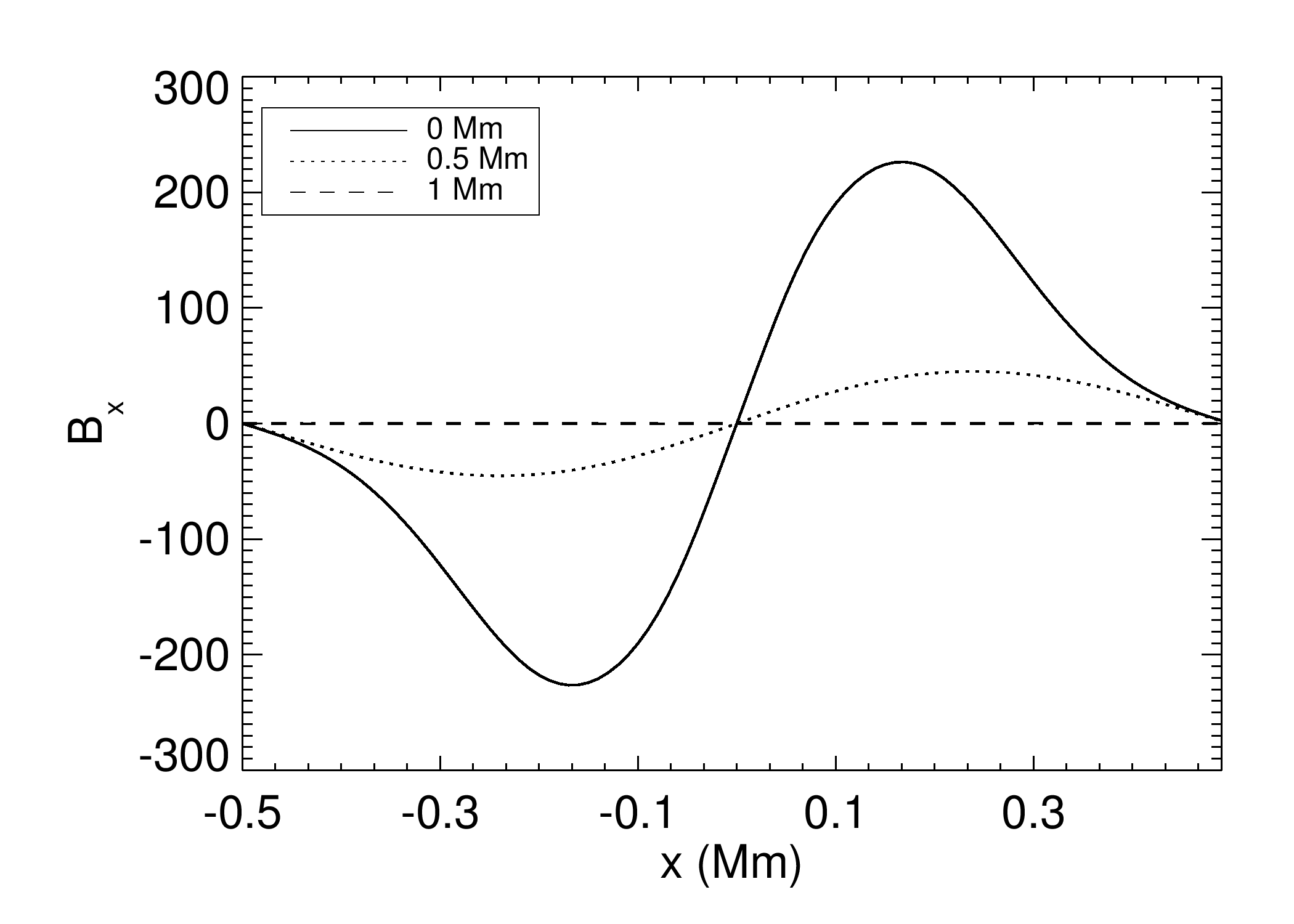}\\
\includegraphics[width=0.95\columnwidth]{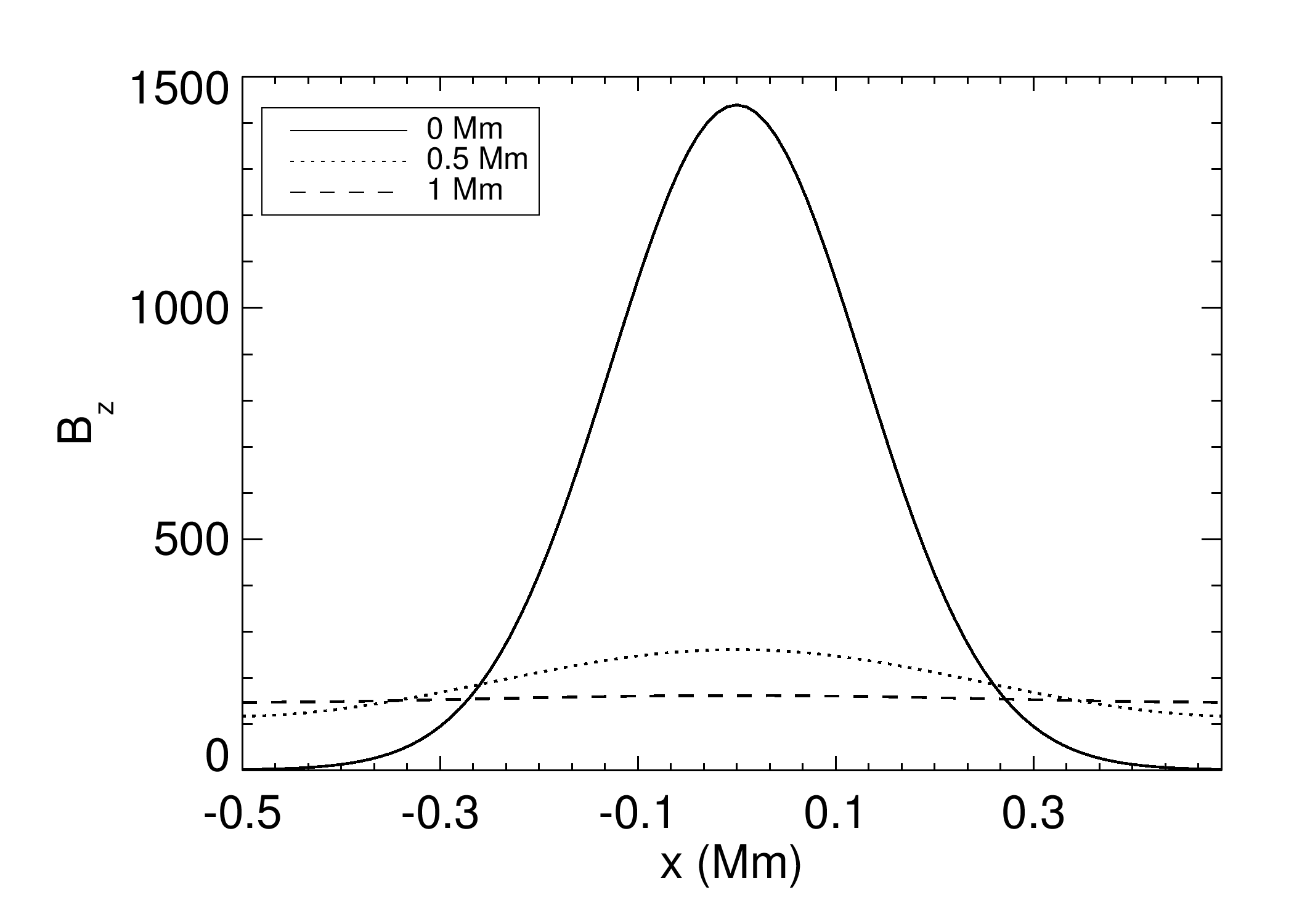}\\
\caption{B$_x$ and B$_z$ components of the magnetic field as a function of the horizontal distance at the following heights: {$z=0$~km (solid curve)}, {$z=0.5$~Mm (dotted curve)}, and {$z=1$~Mm (dashed curve)} of flux tube.}
\label{fig:field_components}
\end{figure}

We model a strong, axially symmetric magnetic flux tube in a stratified solar model atmosphere whose lower boundary is located at photospheric levels.
The horizontal distribution of the magnetic field components (B$_{x}$ and B$_{y}$) across the flux tube at different leveles are shown in Fig.~\ref{fig:field_components}. The initial magnetic field strength on the axis at $z=0$~km is 1435~G, which decreases to 161~G at the top of the box almost to a value close to the magnetic field strength in the ambient medium. The Alfv\'{e}n speed ($v_{\textit{A}}$) in the ambient medium changes by four orders of magnitude as we ascend from $z=0$~Mm to $z=1$~Mm, but it changes only from 10.9~km~s$^{-1}$ to 77.7~km~s$^{-1}$ inside the magnetic flux tube.
The initial model is constructed in a physical domain of size: 1~Mm~x~1~Mm~x~1~Mm and has 100~x~100~x~100 grid points, respectively.
The model considered in our paper is an isolated flux tube which in a Cartesian frame is symmetric with respect to the $x=0$~Mm and $y=0$~Mm planes. The construction of the model is based on a 3D generalisation of the method that was presented in 
\cite{2005ApJ...631.1270H}. 
Close to the base ($z=0$~Mm) the flux tube has a nearly circular cross-section with radius smaller than the half-width of the computational box, $L=0.5$~Mm. For low heights, the flux tube is surrounded by a nearly field-free medium. It expands with height and its radius becomes comparable to $L$. To simulate the effect of neighbouring flux tubes, we assume that the normal component of the magnetic field vanishes at the side walls of the computational domain. More details on the 3-D MHD equilibrium model are given in the Appendix. 
The characteristic physical parameters of the magnetic flux tube are shown in the Table~\ref{tab:properties}. The values in brackets correspond to the ambient medium. The temperature of the magnetic flux tube increases with height, correspondingly increasing the sound speed ($c_{\textit{S}}$) from 7.7~km~s$^{-1}$ to 8.8~km~s$^{-1}$. 
\begin{table*}
\label{tab:photo_equilibrium_table}
\caption{Equilibrium model characteristics for the flux tube.}
\centering
\renewcommand{\arraystretch}{1.5}
\begin{tabular}{lcccccccc}
\hline
Height &  T &  $\rho$ & P &  c$_{\textit{S}}$ &  v$_{\textit{A}}$ & B & $\beta$\\[-1ex]
&{\small (K)} & {\small(kg m$^{\text{-3}}$)} & {\small(N m$^{\text{-2}}$)} & {\small(km s$^{\text{-1}}$)} & {\small(km s$^{\text{-1}}$)} & {\small(G)} & \\
\hline
\multirow{2}{*}{z=1~Mm} & 7263 & 3.4 $\times$ 10$^{\text{-8}}$ & 1.6 & 8.8  & 77.7 & 161 & 0.02 \\[-1ex]
               &{\small  (7195)} &{\small (1.0 $\times$ 10$^{\text{-7}}$)} & {\small (4.8)} & {\small (8.7)} & {\small (39.2)} & {\small (141)} & {\small (0.06)}\\
\multirow{2}{*}{z=0~Mm} & 4768 & 1.3 $\times$ 10$^{\text{-4}}$ & 4.2 $\times$ 10$^{\text{3}}$ & 7.1 & 10.9 & 1435 & 0.5\\[-1ex]
               &{\small  (4766)} &{\small (4.0 $\times$ 10$^{\text{-4}}$)} & {\small (1.2 $\times$ 10$^{\text{4}}$)} & {\small (7.1)} & {\small (0.003)} & {\small (0.77)} & {\small (5.1 $\times$ 10$^{6}$)}\\
                                       \hline
                                      \end{tabular}
					\label{tab:properties}
                                    \end{table*}

\begin{figure}
\includegraphics[width=0.95\columnwidth]{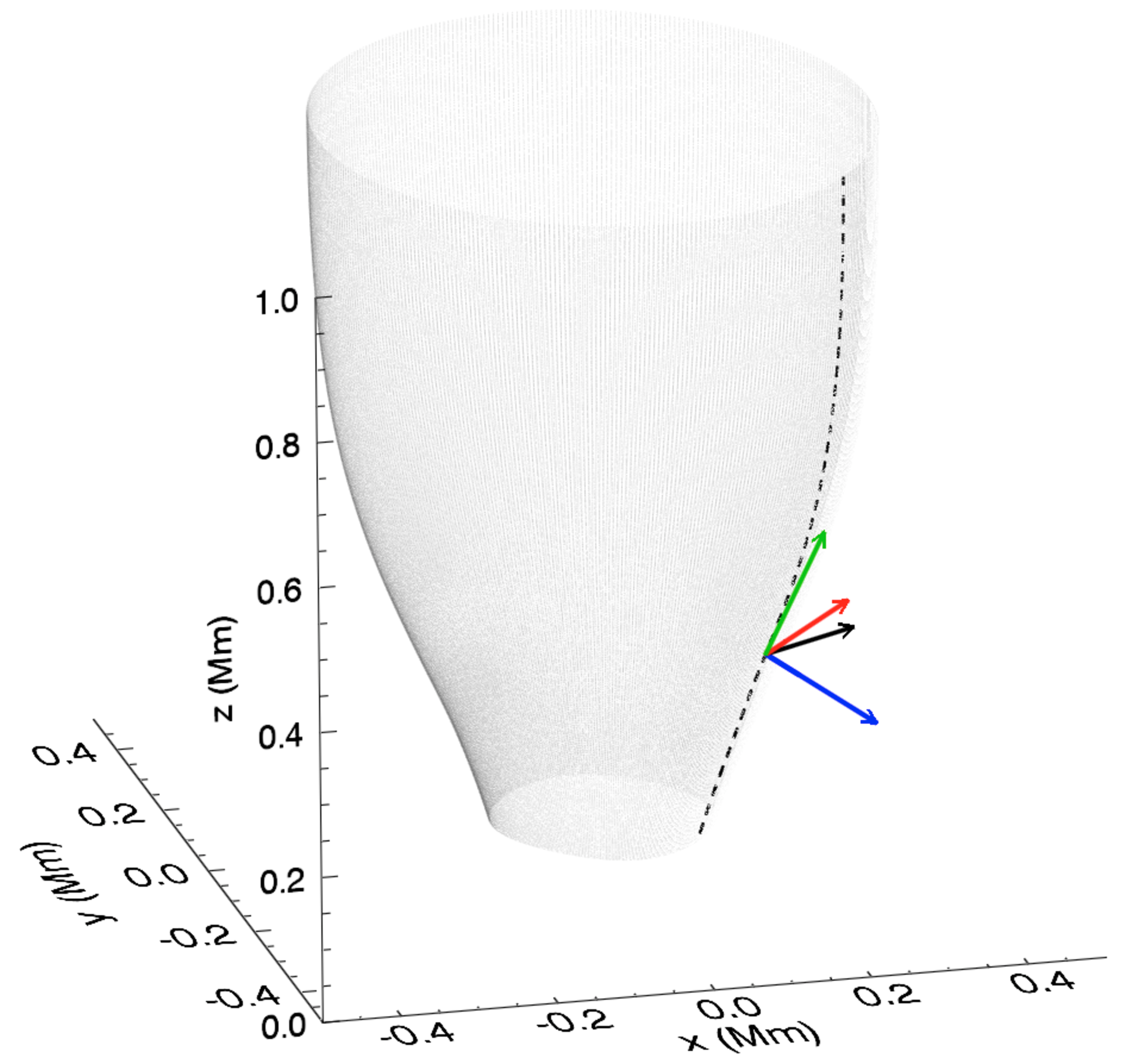}\\
\caption{Three-dimensional representation of the magnetic flux tube. The magnetic flux iso-surface is marked by the field lines (thin gray lines) passing through points at a distance of $r=$0.35~Mm from the axis at a height of $z=$1~Mm. The thick dashed line represents a single field line on this iso-surface.}
\label{fig:fluxtube}
\end{figure}

\begin{figure}
\includegraphics[width=0.95\columnwidth]{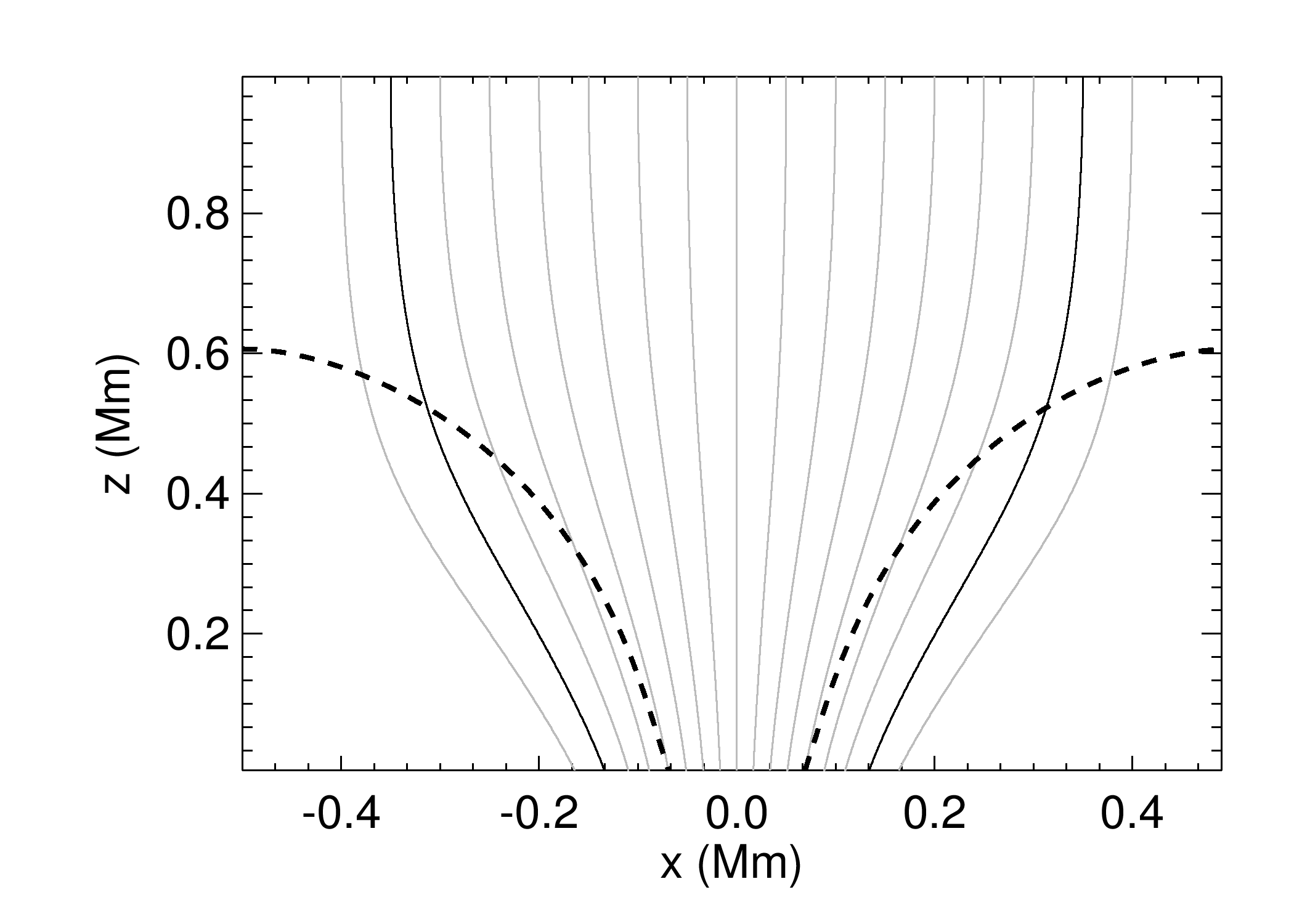}\\
\caption{Two-dimensional representation of the flux-tube. The thin lines mark the magnetic field lines on a $x-z$ plane at $y=$0~Mm. The field lines that trace the iso-surface shown in Fig.\ref{fig:fluxtube} is shown as thick lines. The $\beta=$1 contour is shown as thick dashed line.}
\label{fig:betafieldline}
\end{figure}

To aide the reader in visualizing the model, we plot a three-dimensional representation of the magnetic flux tube in Fig.~\ref{fig:fluxtube} by visualizing field lines that pass through points at distance of $r=0.35$~Mm from the axis and height of $z=1$~Mm. The light gray lines that mark the field lines can be thought as curves that form the magnetic flux iso-surface, since the flux tube that we construct here is axisymmetric. In the figure we also highlight a specific field line (thick dashed curve) and the unit vectors corresponding to different directions at an arbitrary point P on this field line. Further details will be discussed in Sect.~\ref{s:results} where we introduce the various components of velocity with respect to the flux tube. In order to have a clear picture of the shape of the tube and the surface of equipartition between thermal energy density and magnetic energy density, we plot in Figure~\ref{fig:betafieldline} the magnetic field lines of the flux tube on an $x-z$ plane at $y=0$~Mm. The thick curve traces the plasma $\beta = 1$ surface ($\beta$ is the ratio of gas to magnetic pressure) which represents the equipartition level. When viewed in three dimensions, the angle between the magnetic vector and the $\beta=1$ surface normal varies between $0^{\circ}$ and $90^{\circ}$ in the entire computational domain.

\begin{figure}
\includegraphics[width=0.95\columnwidth]{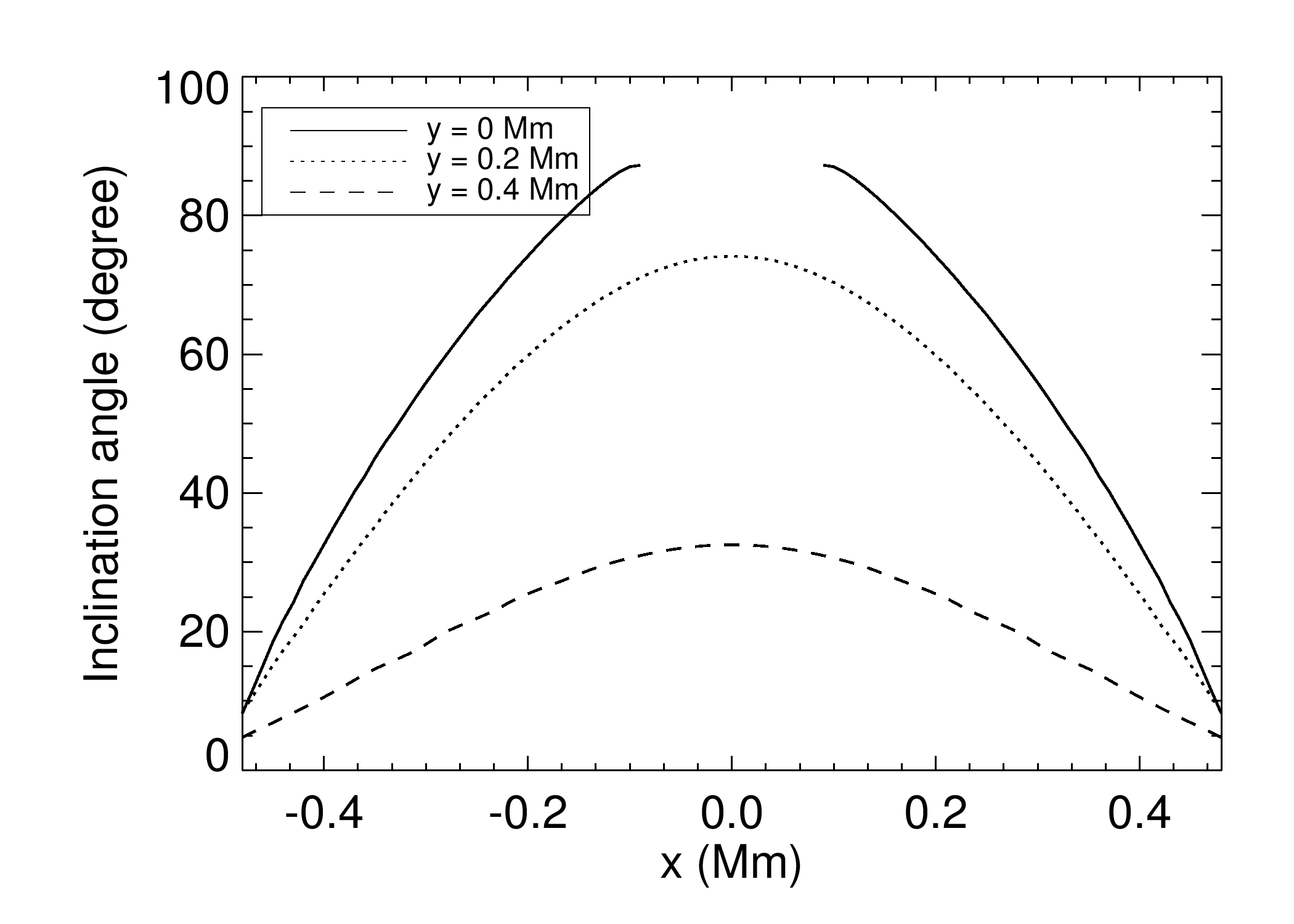}\\
\caption{The inclination of the $\beta=$1 surface normal to the local magnetic field vector. The thick curve is for a plane at $y=$0~Mm. The dotted line is for a plane at $y=$0.2~Mm and the dashed line is for $y=$0.4~Mm.}
\label{fig:betafield_inclination}
\end{figure}

\begin{figure}
\includegraphics[width=0.95\columnwidth]{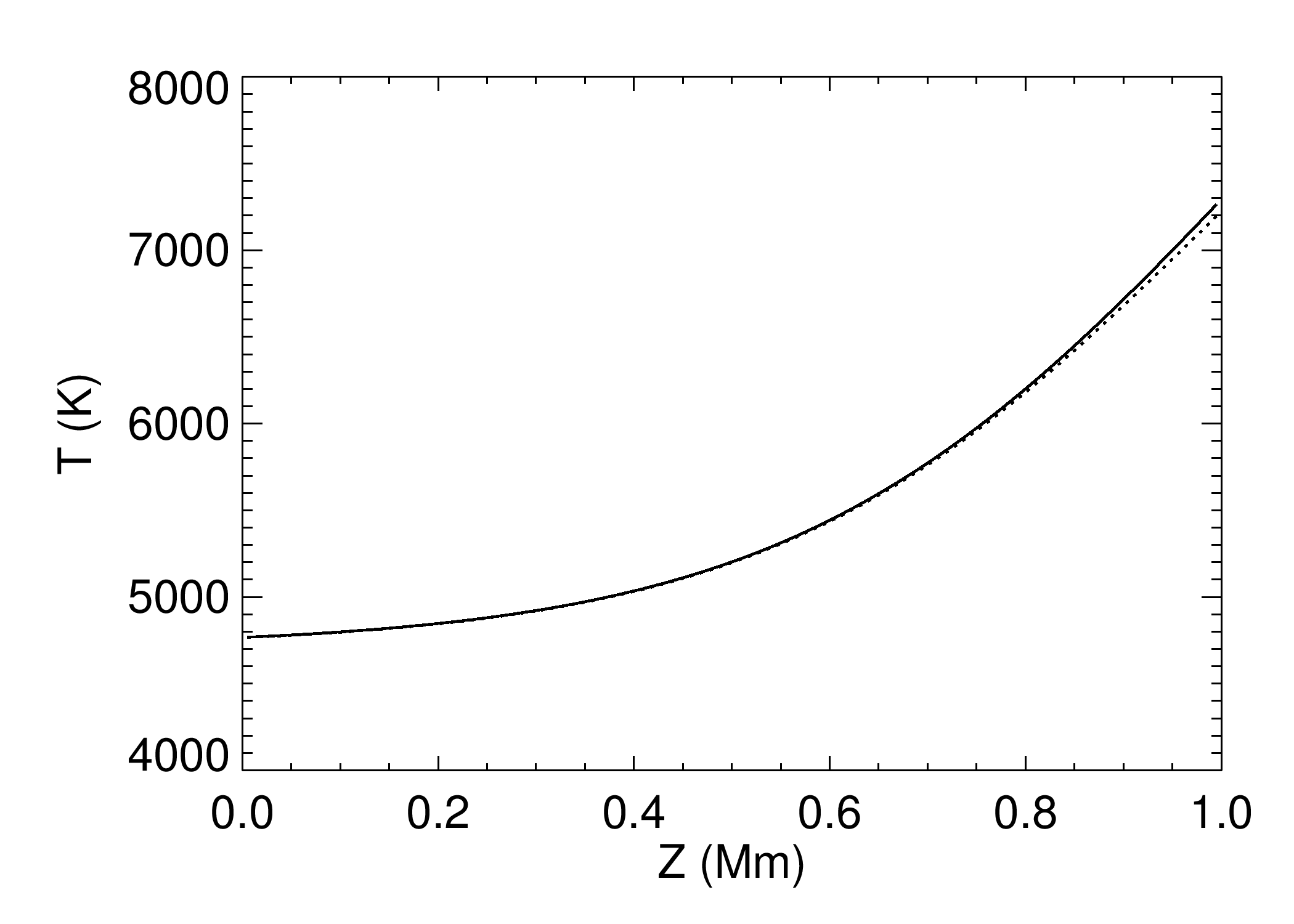}\\
\caption{Temparature as a function of height. The thick curve represents is for the axis and the dotted curve is for the ambient medium.}
\label{fig:plot_temperature_z}
\end{figure}

In Fig.~\ref{fig:betafield_inclination} we plot the inclination angle of the magnetic field vector and the $\beta=1$ surface normal at three different $x-z$ planes corresponding to $y=0$~Mm, 0.2~Mm, and 0.4~Mm. The temperature profile within the flux tube and in the ambient medium of the equilibrium model is shown in Fig.~\ref{fig:plot_temperature_z}.

A typical magnetic flux tube model like the one described above can support the well known three MHD modes. They are the \textit{slow}, \textit{fast} magneto-acoustic and the \textit{intermediate} modes, named so according to the speed of propagation respective to the local Alfv\'{e}n speed  ($v_{\textit{A}}$). In a low-$\beta$ medium, the \textit{slow} wave has a phase speed less than or equal to the sound speed, since the wave is essentially driven as a result of gas pressure perturbations. 
Hodographs or Friedrich diagrams for MHD waves show that the phase speed of the slow wave equals to the sound speed (low-$\beta$) for a direction along the field line and drops to zero as a function of the angle between the propagation vector and the magnetic field vector. On the contrary, the fast wave propagates independent to the field line with a maximum speed equal to $v_{F}$, the fast speed.

\begin{figure}
\centering
\includegraphics[width=0.99\columnwidth]{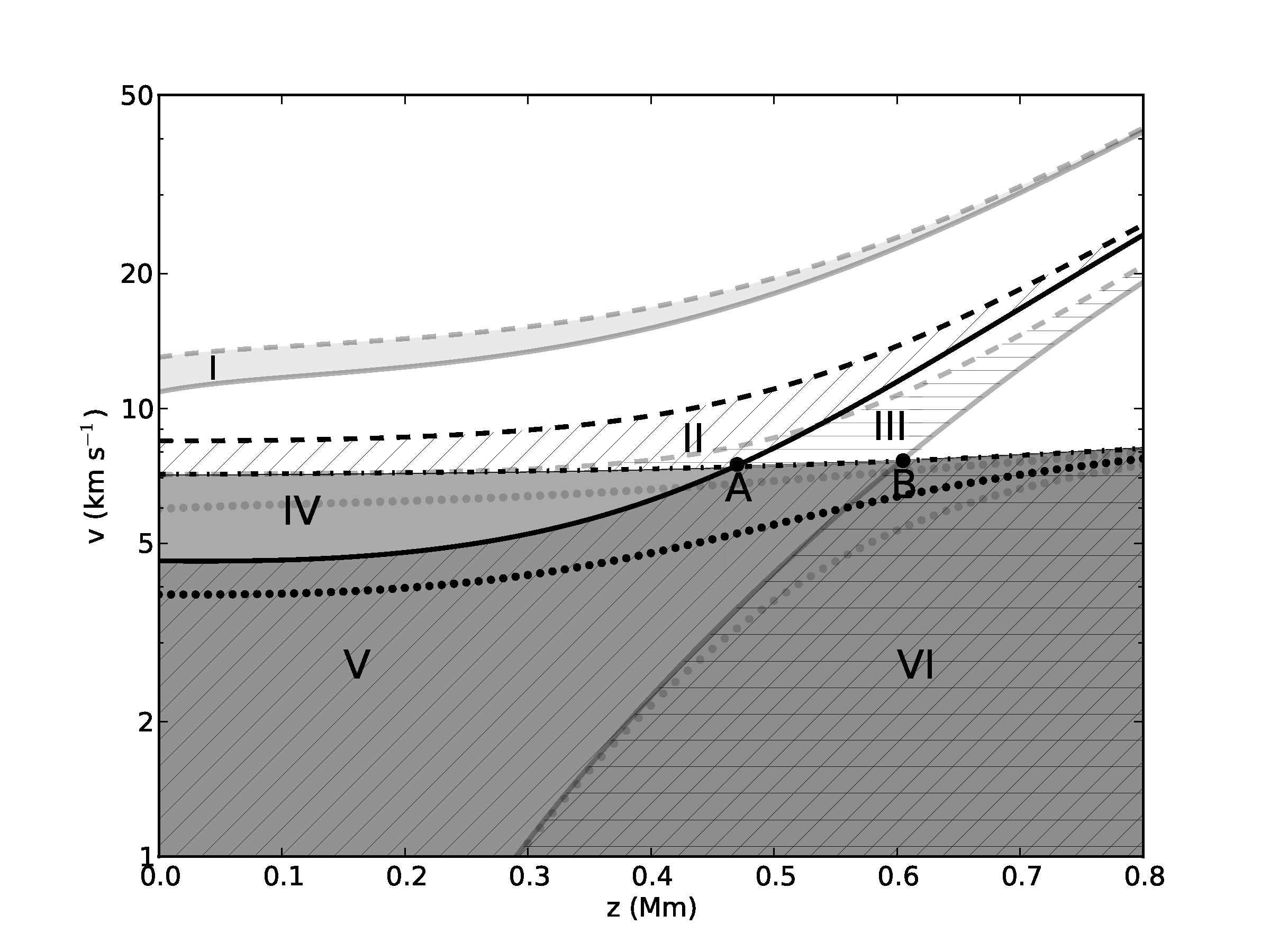}
\caption{Different propagation regimes of the magneto-acoustic waves present in the flux tube as function of height. The plot shows the characteristic speed of waves on the axis of the flux tube (shaded grey: Region I, IV, V \& VI), on the representative field line (slanted line patterns: Region I, V \& VI) and in the ambient medium (horizontal line patterns: Region III \& VI). The characteristic speeds as a function of height are also shown: Alfv\'{e}n (solid), fast (dashed), tube (dotted) and the sound speed (dot-dashed). The thick curves correspond to the values on the representative field line.}
\label{fig:speeds}
\end{figure}

Figure~\ref{fig:speeds} shows the Alfv\'{e}n speed ($v_{\textit{A}}$), sound speed ($c_{\textit{S}}$) and the fast speed ($v_{F}$) defined by:
\begin{equation}
v_{F} = \sqrt{c_{\textit{S}}^{2} + v_{\textit{A}}^2},
\end{equation}
and the tube speed ($c_{T}$) given by:
\begin{equation}
c_{T} = \sqrt{\frac{{c_{S}}^2 {v_{A}}^2}{{c_{S}}^2 + {v_{A}}^2}},
\end{equation}
as a function of height on the axis of the flux tube, on the representative field line and in the ambient medium of the model. 
This plot helps us to identify the different regimes of magneto-acoustic wave propagation in the model. To clearly understand the wave behavior at different locations in the domain, we separate it into six different regions. In the following, $\theta$ represents the angle between the propagation direction and magnetic field vector.\\
\textit{On axis}: The light-grey shaded region (Region I) corresponds to the propagation of fast wave on the axis of the flux tube. The upper limit of propagation speed approaches $v_{F}$ in the direction normal to the field line ($\theta = 90^{\circ}$) and drops to the Alfv\'{e}n speed when $\theta=0^{\circ}$. But, as a function of height, these two speeds become equal and the fast wave speed approaches the Alfv\'{e}n speed. The dark-grey shaded regions (Regions IV, V \& VI) correspond to the propagation of slow waves on the axis of the flux tube. The slow wave has a maximum propagation speed when $\theta=0^{\circ}$ and in this region it is equal to the sound speed. The propagation speed is slower than sound speed in any other direction ($\theta>0^{\circ}$) and vanishes to zero in the direction normal to the field. Hence, in the case of the propagation along the axis of the flux tube, we notice a distinct separation between propagation of slow and fast waves and also that there is a region where there is no wave propagation possible. This scenario remains the same even as one moves away from the axis of the flux tube.\\
\textit{Ambient medium}: Let us now look at the propagation in the ambient medium, represented here by regions III and VI (shown in Fig.~\ref{fig:speeds} by vertical-line pattern). Region III is the propagation zone of the fast wave in the ambient medium. The speed in the direction normal to the field line is $v_{F}$ which is close to sound speed in the lower part of the medium but increases with height as the Alfv\'{e}n speed increases. The propagation at $\theta=0^{\circ}$ is set by the sound speed below the height marked as $B$ at which point the sound speed and the Alfv\'{e}n speed are equal and from there upwards the propagation in the direction of the field is set by the Alfv\'{e}n speed. The slow wave propagation in the ambient medium is marked as dark-shaded region with horizontal strips (Region VI). Up to the height $B$, the Alfv\'{e}n speed (solid grey line at the bottom left part of the figure) is lower than the sound speed and hence the speed of propagation along the field line (the maximum) is the Alfv\'{e}n speed but above the height $B$ it shifts to sound speed. In essence, we do not expect any wave propagation in the ambient medium corresponding to regions IV and V. The fast and slow waves are distinctly separate below the height of point $B$.\\
\textit{Along field line}: Let us analyze the propagation speeds on a field line that crosses the height $z=1$~Mm at radial distance of $r=0.35$~Mm from the axis and azimuth of 1.75$\pi$ radians. The propagation region of the fast wave on this field line is marked in Fig.~\ref{fig:speeds} as Region II and the slow modes are present in Regions V \& VI. Unlike in the ambient medium, the fast wave propagation normal to the magnetic field (equal to $v_{F}$) is larger and we should start seeing an expanding wavefront with height above 0.3 Mm. The propagation speed in the direction of the field remains $c_{S}$ (thick dot-dashed curve) till point $A$ from whereon the propagation speed is $v_{A}$(thick solid curve). The slow waves on this field line exist in regions V and VI (marked by grey-shade and slanted lines). Up to a height of point $A$, the maximum speed of propagation is Alfv\'{e}n speed but it changes to sound speed as the perturbations crosses this height. Similar to the above mentioned cases of propagation along the axis and ambient medium, there exists a region where we do not expect any wave propagation (Region IV). Due to the increasing Alfv\'{e}n speed with height, the speed of fast mode is approximately equal to the Alfv\'{e}n speed on the flux tube axis and hence distinguishing between the \textit{intermediate} Alfv\'{e}n mode and the \textit{fast} mode is in practice rather difficult, unless polarization information is available. In the ambient medium 
the difference between $v_{F}$ and $v_{A}$ is slightly larger, making it feasible to separate out the three modes. The excitation mechanisms, that we consider in this paper, generates magneto-acoustic waves within the magnetic flux tube as a result of the driving motion at the footpoint. Since we do not have a variation within the magnetic flux tube from high-$\beta$ to low-$\beta$ plasma, we cannot separate the energy contribution from the fast and Alfv\'{e}n modes.

\section{Wave Excitation}\label{s:wave_excitation}

The study of the excitation of waves in a vertically oriented magnetic flux tube embedded in an otherwise unmagnetized ambient medium has been discussed in a series of papers by 
\citeauthor{2002ApJ...573..418M}
\citep{1989ApJ...337..470M,1995ApJ...448..865M,2000ApJ...541..410M, 2002ApJ...573..418M}. 
They have outlined the theory of the interaction of a flux tube with turbulent convection and the consequent generation of waves in the magnetic feature for solar and in general for stellar atmospheres. Excitation by granular buffetting, a process by which waves can be generated in magnetic flux tubes have been studied by
\citeauthor{2000ApJ...535L..67H}
for a flux tube in a photospheric network
\citep{1999ApJ...519..899H,2000ApJ...535L..67H},
and for explaining the heating of the chromospheric network
\citep{2005ApJ...631.1270H,2008ApJ...680.1542H}.

In a way similar to
\cite{2005ApJ...631.1270H} and others
\citep{2008ApJ...680.1542H,2009A&A...508..951V,2011AnGeo..29.1029F}, we investigate wave propagation by a range of  excitation in the equilibrium model.
The excitation is driven at the lower boundary by prescribed drivers that model different wave generation  mechanisms at the photospheric layer. Our earlier work
\citep{2009A&A...508..951V,2011SoPh..tmp..349V} 
considered wave excitation in a 2D equilibrium magnetic field configuration by a transverse motion of the lower boundary similar to 
\cite{2005ApJ...631.1270H}. 
The possible wave modes generated in this scenario were restricted to the slow and fast magneto-acoustic modes. Since we now turn to a 3D configuration in this paper, studying the properties of yet another kind of wave, namely, the intermediate Alfv\'{e}n wave is now also possible
\citep{2011AnGeo..29.1029F}.
Nevertheless, it would be worthwhile to look into the 3D problem of granular buffetting, which is not likely to generate any Alfv\'{e}n waves.

\subsection{Horizontal driver}\label{ss:hor_wave_driver}
We implement a horizontal driver to mimic the granular buffetting motion as observed in photospheric layers. The transverse driving velocity at the bottom boundary is specified as follows:
\begin{eqnarray*}
                                          V_{x}(x,y,0,t) = \left\{ \begin{array}{l l} 
                                            \displaystyle V_{0} \sin \left(\frac{2 \pi t}{P}\right) 
                                            &\quad 0 \le t \le P/2,\, \\[1ex]
                                            \displaystyle 0
                                            &\quad \phantom{0 >  }t > P/2.\,
                                          \end{array} \right.
                                          \label{eq:horizontal_velocity}
                                        \end{eqnarray*}
The amplitude of the motion ($V_{\text{0}}$) is 750~m~s$^{\text{-1}}$ and wave period ($P$) is 24~s which is quantitatively in the range of values found in subsurface convective layers in which the flux tube is rooted
\citep{1996ApJ...463..365B,2010A&A...511A..39U,2011ApJ...740L..40K}.

\subsection{Torsional driver}\label{ss:tor_wave_driver}

In order to study the effect of vortex-like motion in a flux tube, we also use a torsional driver. In this case, the azimuthal component of the velocity is specified as follows:
					\begin{eqnarray*}
                                          V_{\phi}(x,y,0,t) = \left\{ \begin{array}{l l} 
                                            \displaystyle -V_{0} \tanh \left(\frac{2 \pi r}{\delta r}\right) \sin\left(\frac{2 \pi t}{P}\right)
                                            &\quad 0 \le t \le P/2,\, \\[1ex]
                                            \displaystyle 0
                                            &\quad \phantom{0 >  } t > P/2.\,
                                          \end{array} \right.
                                          \label{eq:torsional_velocity}
                                      \end{eqnarray*}
Here $\delta r$ describes the characteristic distance by which the azimuthal component changes from 0 to $V_{0}$. For a practical purpose we use $\delta r=1$~Mm. The amplitude and period in this case are the same as for the horizontal driver.

The driving motion prescribed above generates a wide range of wave-like disturbances in the medium which we analyze in the following section.

\section{Numerical Methods and Boundary Conditions}\label{s:numerical_methods}
The 3D numerical simulations were carried out using the Sheffield Advanced Code (SAC)
\citep{2008A&A...486..655S}.
SAC is a fully non-linear MHD code based on a more general-purpose code named VAC 
\citep{1996ApL&C..34..245T}. SAC uses a modified version of the set of MHD equations to deal with a strongly-stratified magnetized medium. 
The spatial derivatives are calculated using a fourth-order central difference scheme and the time derivatives are calculated using a Runge-Kutta scheme.
Various boundary conditions are available for the different physical problems. For our purpose we use open boundary conditions for the top and the side boundaries. This allows any generated perturbation to propagate out of the simulation domain without reflection.

\section{Simulation Results}\label{s:results}

Here we have considered a strong magnetic flux tube with the plasma-$\beta$=1 surface outlining the boundary of the constructed magnetic flux tube. Since, given that, inside the magnetic flux tube plasma $\beta < 1$, the magneto-acoustic waves generated by the driver are the slow (predominantly acoustic) and fast (predominantly magnetic) waves
\citep[see][]{2009A&A...508..951V}.   

\begin{figure}[h]
\centering
\includegraphics[width=0.95\columnwidth]{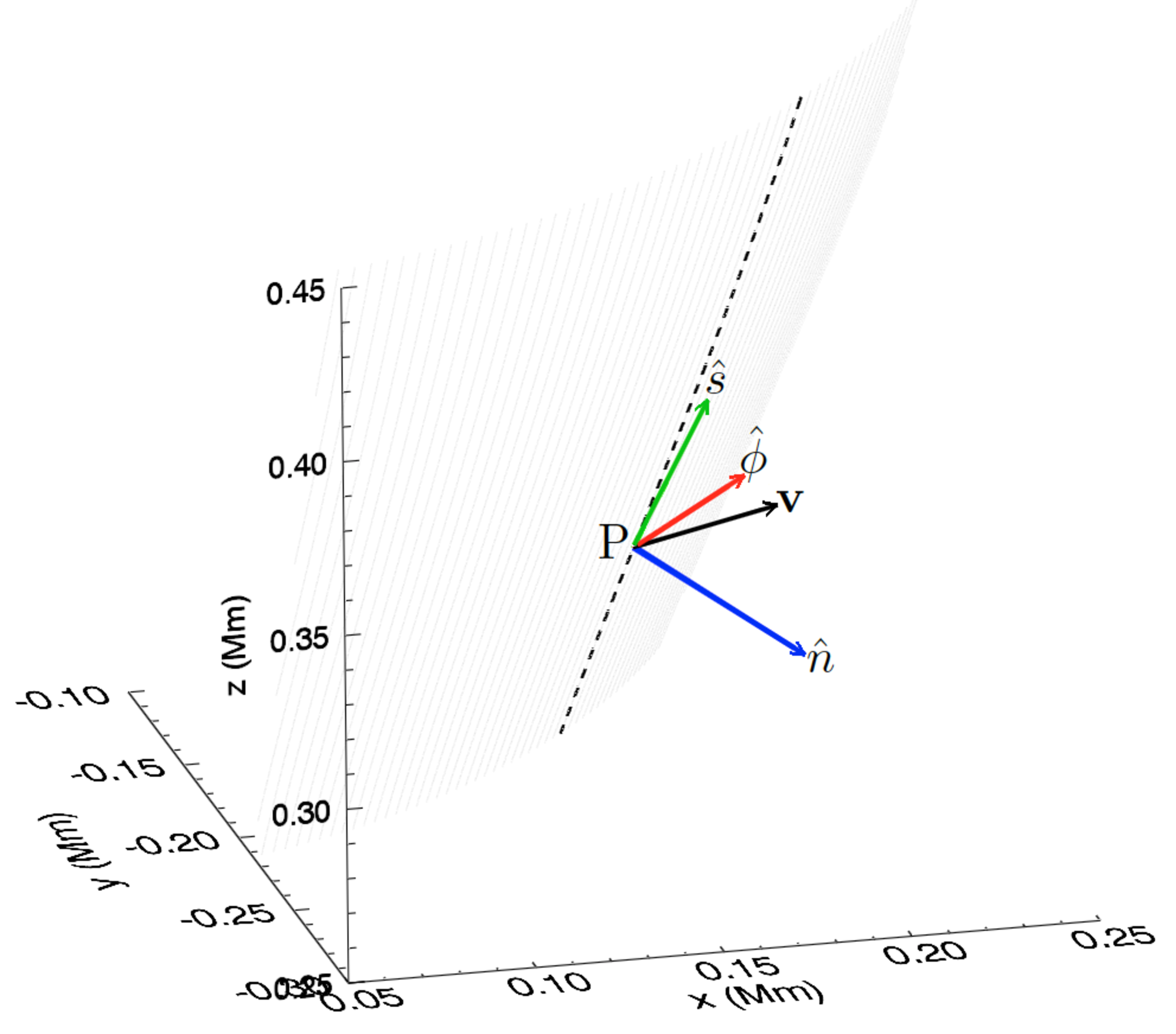}
\caption{The description of the parallel ($v_{\parallel}$), perpendicular ($v_{\perp}$) and the azimuthal ($v_{\phi}$) components of velocity vectors with respect to the magnetic field line and the equipotential surface.}
\label{fig:geometry}
\end{figure}

The gas pressure perturbations are directed along the field lines. The slow magneto-acoustic wave (SMAW) in a low-$\beta$ plasma can be visualized by the parallel component of velocity ($v_{\parallel}$) to the background magnetic field, as they are predominantly acoustic in nature. On the other hand, as the fast magneto-acoustic wave (FMAW) can also propagate across the magnetic field lines, it can be identified in the perpendicular components of velocity ($v_{\perp}$) to the background magnetic lines of force. The propagation of the intermediate mode is characterized by the azimuthal component ($v_{\phi}$) of the velocity. The description of the three components of velocity is shown in Figure~\ref{fig:geometry}. 

In the following, we describe the method used to compute the three components of velocity. Since the initial model is axisymmetric, we can assume that the cross-section of the iso-surface, or the surface of equal magnetic potential of the flux tube at a given height traces a circle with the center at axis. See Fig.~\ref{fig:fluxtube} for more information on the geometrical shape of the flux tube and the position of the field line. We proceed by calculating the field lines that pass through these points at time $t=0$. The displacement of the starting points are determined at each timestep and the new magnetic field lines are traced as they evolve. The driving motion at the base distorts the flux tube and consequently the flux tube departs from being axisymmetric. However, the field lines that define the magnetic flux iso-surface remain the same. For brevity, we consider a single field line at a radius of r=0.35~Mm and azimuthal angle of $\phi=1.75\pi$ radians (dashed curve in Fig.~\ref{fig:fluxtube} \& Fig.~\ref{fig:geometry}). The parallel component of velocity ($v_{\parallel}$) along the field line ($\hat{s}$) can be computed by taking the scalar product of $\textbf{v}$ on $\textbf{B}$.  We construct the vector normal to the surface of equal magnetic potential ($\hat{n}$) by using two adjacent field lines on the points that form the circle. The set of points which define the two adjacent field lines form a plane and we use these points to calculate the surface normal, $\hat{n}$. The projection of the velocity vector (\textbf{v}) on this vector gives the normal component. The cross product of the magnetic field vector (\textbf{B}) and the normal vector then gives the vector tangent to the magnetic equipotential surface ($\mathrm{\hat{\phi}}$). The projection of the velocity (\textbf{v}) on this vector gives the azimuthal component, $v_{\mathrm{\phi}}$. The parallel ($\hat{s}$), normal ($\hat{n}$) and the azimuthal ($\hat{\phi}$) vectors for a point $P$ on the selected field line (thick dashed line) are shown in Fig.~\ref{fig:geometry}. The magnetic field lines that form the iso-surface are shown as set of thin lines. Determining the velocity components will help in separating out the contribution of different modes that are generated as a result of the driving motion at the lower boundary.\\

In the following we discuss the result of our numerical simulation for the two drivers discussed in the previous section.

\subsection{Horizontal excitation}\label{ss:hor_wave_dynamics}
\begin{figure*}
\centering
\includegraphics[width=0.9\textwidth]{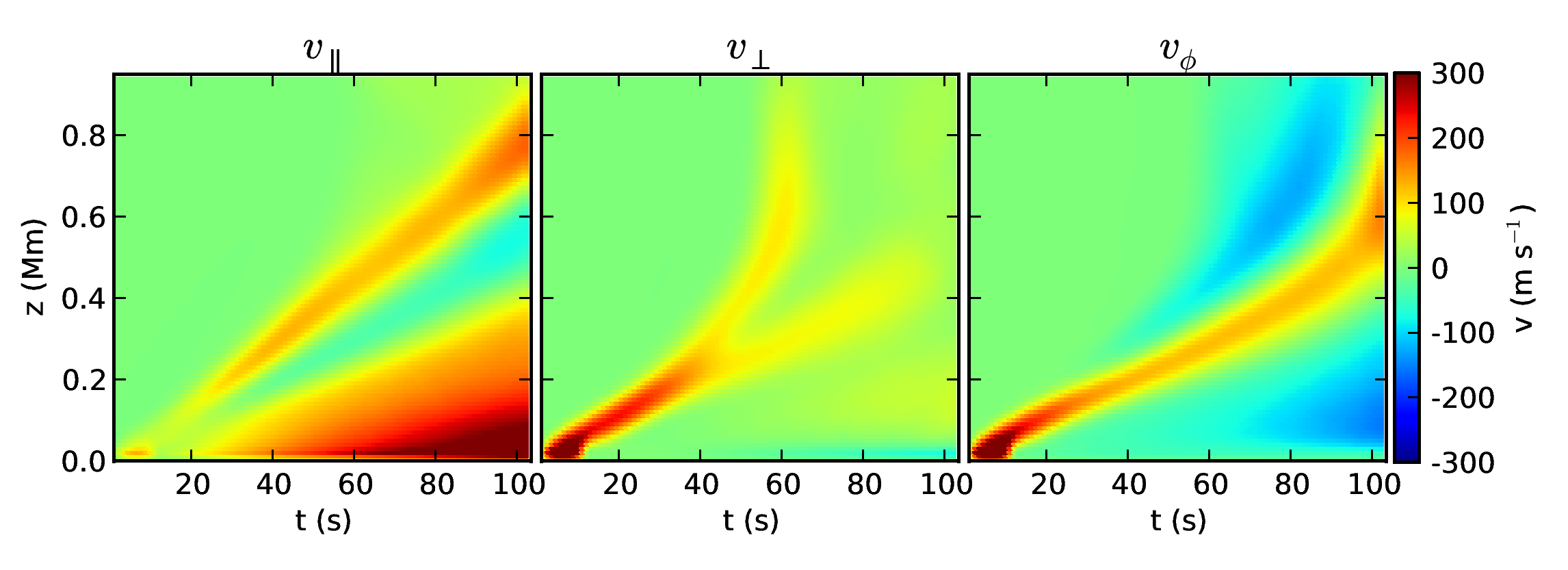}
\caption{The time distance plot of the three components of velocity ($v_\parallel$,$v_{\perp}$ and $v_{\phi}$) along the vertical path slightly away from the axis of the magnetic flux tube as a result of horizontal excitation.}
\label{fig:hor_td}
\end{figure*}

We consider the case in which the magnetic flux tube is excited by a uni-directional horizontal motion at the lower boundary. The physical scenario is similar to the case wherein a deep rooted flux tube in the interganular lane is buffeted by surrounding granules. From the dynamics of the magnetic bright points (MBPs), as studied by
\cite{2005ApJS..156..265C}, 
this motion corresponds to a single instance wherein the magnetic fluxtube receives a random kick from a direction not necessarily the same along which it was already moving.
Such a transversal driver generates strong SMAW and FMAW in the medium that propagate along the magnetic flux tube. 
Figure~\ref{fig:hor_td} show the time-distance plots of the velocity component as described in the beginning of this section. These panels correspond to $v_{\parallel}$, $v_{\perp}$ and $v_{\phi}$ along a vertical path parallel to and about 140~km away from the magnetic flux tube axis. 

The longitudinal perturbations produced by the passage of the slow wave along the field line can be visualized by plotting the parallel component of velocity ($v_{\parallel}$) on the flux tube magnetic field. We find dominant velocity perturbations associated with the SMAW. This driver also generates relevant FMAW in the medium. Since the FMAW can also propagate across the magnetic field lines, we determine the perpendicular component of velocity ($v_{\perp}$) to the magnetic field of the flux tube. 
The azimuthal component of velocity ($v_{\phi}$) has a comparable magnitude, as seen from the contour values. This is a result of the location of the particular field line under consideration with respect to the driving motion which results in a significant contribution to the azimuthal component.

\begin{figure*}
\centering
\includegraphics[width=0.9\columnwidth]{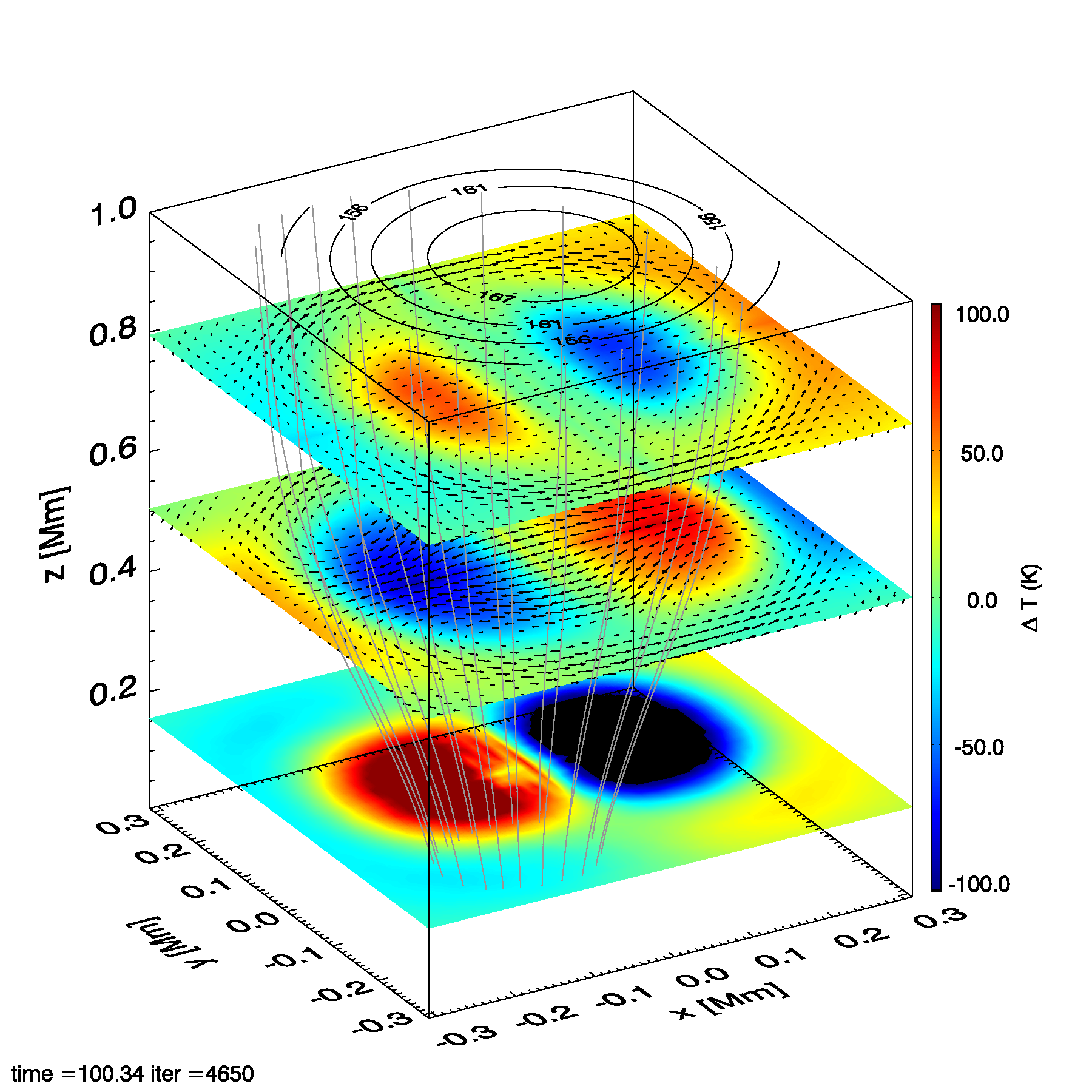}
\includegraphics[width=0.9\columnwidth]{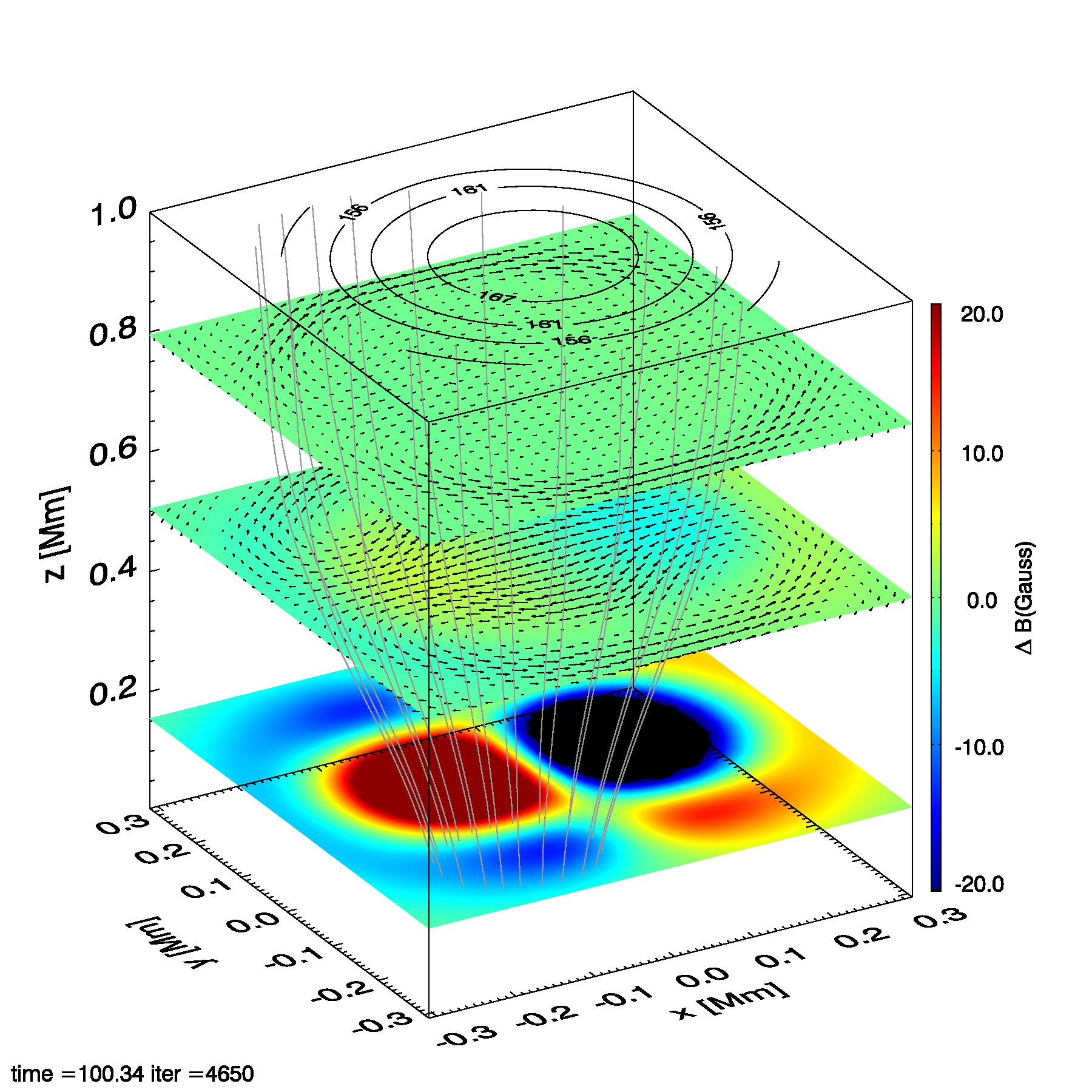}
\caption{Temperature (left) and magnetic field strength (right) fluctuations in a flux tube at three different heights ($z=$150~km, 500~km and 800~km) at t=100~s as result of a transverse uni-directional excitation. Field lines are shown to mark the flux tube. The contour of equal magnetic field strength at $z=$800~km are shown on the top.}
\label{fig:hor_snapshot}
\end{figure*}

Figure~\ref{fig:hor_snapshot} shows a snapshot (at t=100~s) of the temperature (left panel) and the magnetic field (right panel) fluctuations on $x-y$ planes at three different heights ($z$=150~km, 500~km and 800~km). Few field lines mark the magnetic flux tube and the variation of the field strength along the field lines are color coded. The snapshot depicts a clear signature of the passage of the SMAW at height $z=$500~km. Glancing through the sequence of the snapshots, in temporal order:
\begin{itemize}
\item {at t$\sim$35~s, strong velocity and the magnetic field perturbations suggest the arrival of the fast MHD mode at $z=$500~km;}
\item {at t$\sim$46~s, the fast MHD mode crosses at $z=$800~km level resulting in amplified velocities and magnetic field perturbations. The circular contours corresponding to a constant magnetic field strength values, shown on top of the box, shifted with respect to the time at which the FMAW reaches the top of the box;}
\item {at t$\sim$67~s, the slow MHD mode crosses at $z=$500~km causing strong temperature fluctuations ($\delta T$);}
\item {similarly at, $t=$100~s (Fig.~\ref{fig:hor_snapshot}), there is a strong evidence of $\delta T$ which is absent in the case of a torsional excitation.}
\end{itemize}

Since the flux tube has a substantial $B_{x}$ component at the footpoint as seen in Fig.~\ref{fig:field_components}, the velocity of the uni-directional driver has a larger and more dominant component parallel to the field in a small region on either side of the magnetic flux tube. We see strong $v_{\parallel}$ components on either side of the magnetic flux tube concentrated on the axis associated with the SMAW.

\subsection{Torsional excitation}\label{ss:tor_wave_dynamics}
Observations of the solar photosphere reveal that bright points corresponding to flux tubes located in the intergranular lane may follow a spiral path centered at regions where boundaries of two or more granules meet
\citep{2008ApJ...687L.131B}. 
One assumes that an effective torque acts on these flux tubes, and they start to rotate as they follow such a trajectory. The azimuthal component becomes amplified, as these flux tubes approach the strong down-flow regions at the confluence of two or three intergranular lanes. Assuming that the simulated magnetic flux tube is located where the azimuthal driving is stronger, we study the dominant wave modes that are likely to be excited.

\begin{figure*}
\centering
\includegraphics[width=0.9\textwidth]{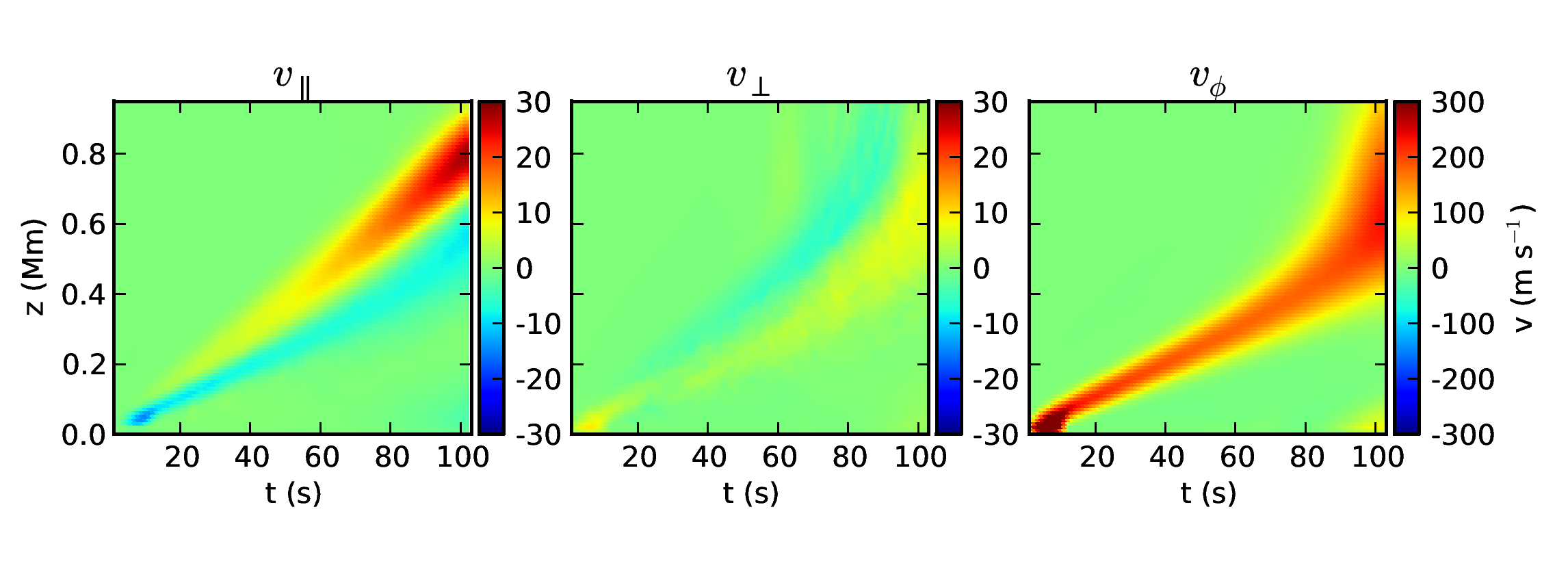}
\caption{The time distance plot of the three components of velocity ($v_\parallel$,$v_{\perp}$ and $v_{\phi}$) slightly away from the axis of the magnetic flux tube as a result of torsional excitation.}
\label{fig:tor_timed}
\end{figure*}

Figure~\ref{fig:tor_timed} shows the time-distance plot of $v_{\parallel}$, $v_{\perp}$ and $v_\phi$ components of velocity where the MHD perturbations are driven by a torsional driver.
The $v_{\parallel}$ amplitude in the case of torsional excitation is rather small ($\sim$30 ms$^{\text{-1}}$) compared to the horizontal excitation ($\sim$300 ms$^{\text{-1}}$), suggesting that the torsional motion generates SMAW that are weaker than those generated by a horizontal driver.
This feature is also seen in the temperature fluctuations, which are mainly associated with the slow MHD mode. Also, the torsional driver generates weaker temperature fluctuations than the horizontal driver. The time-distance plot of $v_{\parallel}$ shows a weak fast MHD mode branch that splits from the dominant slow mode branch which is absent in the case of a horizontal excitation.

We noted above, in the case of horizontal excitation (Sec.~\ref{ss:hor_wave_dynamics}), that the actual values are calculated on a vertical line parallel to the magnetic flux tube axis. Since the magnetic flux tube expands with height, the representative vertical line crosses a range of different magnetic equipotential surfaces, significantly in the lower part of the computational box. The propagation of the intermediate mode is characterized by the rotation of magnetic equipotential surfaces with different amplitudes as one moves away from the flux tube axis. We have found that the magnetic flux tube suffers a torsional kick and this motion is carried away along the field lines. The contours of constant magnetic flux value rotate independently of each other as can be seen at the top of the box. Even though the $B_{x}$ component at the foot point of the magnetic flux tube is significant, the velocity of the torsional driver is in the plane normal to the magnetic flux tube and hence nowhere parallel to the field. Thus, SMAW are not excited directly and dominantly in this case. 

\subsection{Energy transport}\label{ss:energy_transport}

Let us now consider the energy transport by MHD waves for the two driving cases detailed in the previous section. Following
\cite{2009A&A...508..951V},  
instead of using the full non-linear equation for energy flux, we consider the linearized wave-energy flux equation given by
\cite{1974soch.book.....B}.   
The acoustic ($F_{A}$) and Poynting fluxes ($F_{P}$) are computed for the two cases as follows:
\begin{equation}
{\textbf F_{A}}=\Delta p{\textbf V},
\label{eq:aco_equation}
\end{equation}
\begin{equation}
{\textbf F_{P}}=\frac{1}{4 \pi}({\textbf B}_{0} \cdot \Delta
{\textbf B}){\textbf V} - \frac{1}{4 \pi}({\textbf V} \cdot \Delta
{\textbf B}){\textbf B}_{0},
\label{eq:poy_equation}
\end{equation}
where $\Delta$ represents the perturbations with respect to the equilibrium and \textbf{B}$_{0}$ is the equilibrium magnetic field. Comparing the acoustic flux generated by the horizontal and torsional drivers, we find that in the case of a horizontal driver, there is a significant amount of acoustic flux crossing the height $z=$500~km. We already discussed in Sec.~\ref{ss:hor_wave_dynamics}, the uni-directional driver generates strong SMAW. Hence, we see a strong  acoustic flux (longitudinally dominated) associated with it. This is absent in the case of torsional driver, as there is negligible amount of slow wave generation (see Sec.~\ref{ss:tor_wave_dynamics}).

Since the slow, fast and torsional Alfv\'{e}n modes are present in the medium, it is inspiring to calculate the contribution of these three modes in terms of the wave energy that they transport. The fast and slow MHD mode contribute to the acoustic ($F_{A}$) and Poynting fluxes ($F_{P}$) over different scales depending on the plasma-$\beta$. Assuming that the driver implemented here generates linear waves, we have calculated the wave energy fluxes of the different modes for two cases.
For any given mode, the energy of the wave motion transported through the medium depends on the direction with respect to the magnetic field and maximum speed of the mode in the particular direction. Decomposing these energy components in the three directions, the wave energy fluxes can be computed as follows,

\begin{equation}
F_{\parallel} = \rho v_{\parallel}^{2} c_{\textit{S}},
\end{equation}

\begin{equation}
F_{\perp} = \rho v_{\perp}^{2} \sqrt{c_{\textit{S}}^{2} + v_{\textit{A}}^{2}},
\end{equation}

\begin{equation}
F_{\phi} = \rho v_{\phi}^{2} v_{\textit{A}},
\end{equation}
where, $F_\parallel$, $F_{\perp}$ and $F_{\phi}$ are the parallel, perpendicular and azimuthal energy fluxes corresponding to the low plasma-$\beta$ propagation of slow, fast and intermediate wave, respectively. It should be noted that the perpendicular component of the time-averaged energy computed by 
\cite{2012ApJ...746...68K}
is slightly different from the one that is used here. The maximum speed of the FMAW in the perpendicular direction to the magnetic field is equal to $v_{F}$. Although $v_{F}$ is equal to the Alfv\'{e}n  speed at higher layers on the axis of the flux tube, this is not the case for the particular field line that we have chosen as shown in Fig.~\ref{fig:speeds}. Hence, it is appropriate to use $v_{F}$ to calculate the perpendicular wave energy flux.
 
\begin{figure*}
\centering
\includegraphics[width=0.9\textwidth]{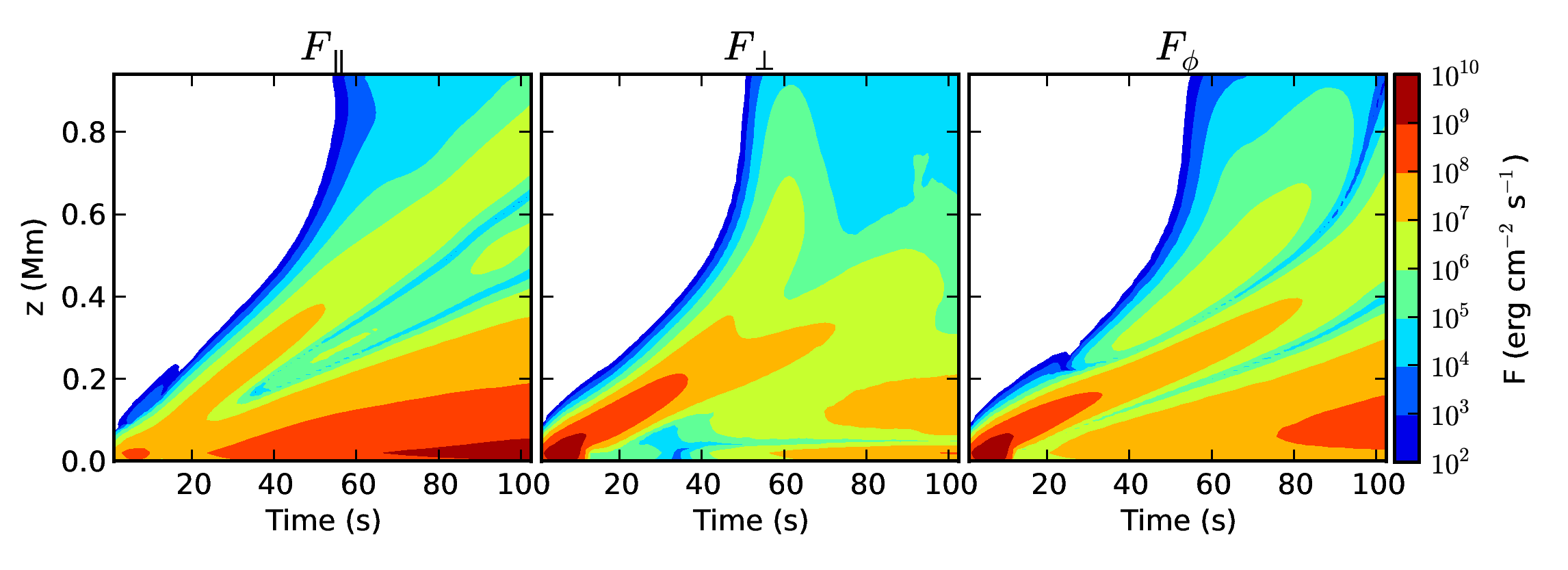}
\caption{The time distance plot of the three energy fluxes ($F_\parallel$,$F_{\perp}$ and $F_{\phi}$), corresponding to the slow, fast and intermediate wave for a distance along the vertical line slightly away from the axis of the magnetic flux tube as a result of horizontal excitation.}
\label{fig:hor_td_flux}
\end{figure*}

\begin{figure*}
\centering
\includegraphics[width=0.9\textwidth]{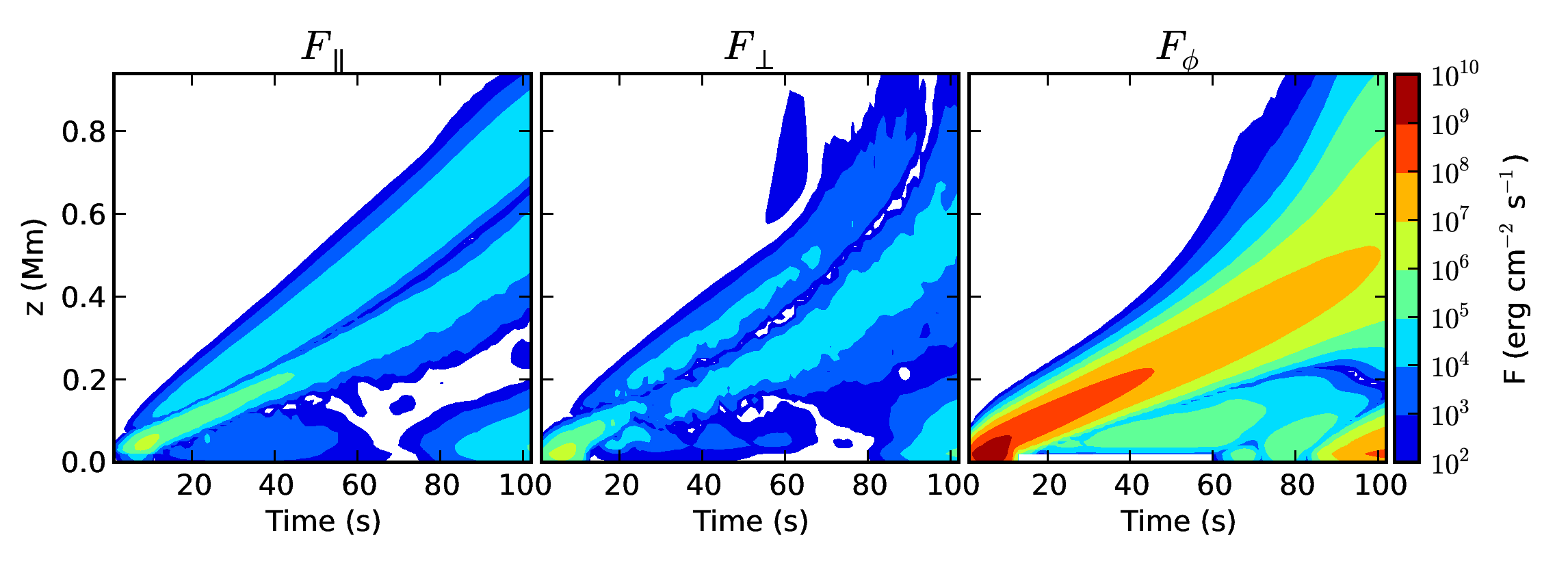}
\caption{The time distance plot of the three energy fluxes ($F_\parallel$,$F_{\perp}$ and $F_{\phi}$), corresponding to the slow, fast and intermediate wave for a distance along the vertical line parallel to the axis of the magnetic flux tube as a result of torsional excitation.}
\label{fig:tor_td_flux}
\end{figure*}

Figures~\ref{fig:hor_td_flux} and \ref{fig:tor_td_flux} show the time-distance plot of the three energy fluxes ($F_\parallel$, $F_{\perp}$ and $F_{\phi}$) computed on the representative magnetic field line as a result of horizontal and torsional excitation in a flux tube, respectively. At larger heights, the parallel flux ($F_\parallel$) corresponding to predominantly SMAW for the case of uni-directionally excited flux tube is $\mathcal O(10^{7}$) erg cm$^{-2}$ s$^{-1}$, which is two orders of magnitude larger than for the case of a torsionally excited flux tube. 
There is a significant amount of  perpendicular flux, $\mathcal O(10^{6}$) erg cm$^{-2}$ s$^{-1}$, being transported close to the Alfv\'{e}n speed in the case of horizontal excitation which is also absent in torsionally excited flux tube. As noted in Sec.~\ref{ss:hor_wave_dynamics}, the contribution to the azimuthal component as a result of the location of the flux tube with respect to the driving motion results in a significant azimuthal energy flux for the unidirectional case. There is indeed a more stronger azimuthal energy flux for the torsionally excited flux tube. 

\section{Summary and Conclusion}\label{s:conclusion}

Three-dimensional modelling of the MHD wave propagation in a magnetic flux tube embedded in a highly stratified solar atmosphere was carried out.
We have investigated two types of excitation mechanisms that are characteristic processes by which waves can be generated in magnetic flux concentrations of the real solar atmosphere.
We considered a transversal driver that mimics granular buffeting and a torsional driver which can be approximated to observed vortex flows within flux tubes.
The driving motion generates slow and fast magneto-acoustic waves in the medium.
The granular buffeting seems to generate strong SMAWs and FMAWs, when compared to vortex-like motion.
The longitudinal acoustic flux transported by these waves also varies depending on the excitation mechanism. We found that, as a result of the granular buffetting simulated here using a unidirectional horizontal motion of the lower boundary, more acoustic energy is generated and this is likely to become dissipated at chromospheric heights. 
Hence, we conclude that the granular buffeting produces stronger emission in the chromospheric network when compared to the vortex flows driven by convective downflow.
The observations of vortex flows in the photosphere and the corresponding chromospheric feature reveal that the bright points associated with the flux concentration follow a spiral path towards the center of the vortex. Although high-resolution observations do not completely resolve these features yet, one may realistically suspect that these bright points are constantly buffetted by the granules on their way to the intergranular sink. We suggest that these buffeting results in a horizontal excitation of the magnetic flux tube, generating slow longitudinally dominant MHD waves and FMAW which eventually dissipate due to shock formation and show up as emission in the chromosphere. However, as these flux tubes approach the eye of the vortex, the horizontal excitation ceases and the flux tube is mainly driven by a torsional motion, which generates negligible SMAWs and FMAWs. This results in lesser shocks in the upper atmosphere and, hence, a weaker emission.

Apart from the other sources of wave driven heating mechanisms like, strong downflows in the outskirts of magnetic elements resulting in excitation of slow modes within the flux tube developing into shocks 
\citep{2011ApJ...730L..24K}, 
transversely shifting magnetic field lines within the magnetic concentration producing Alfv\'{e}n waves causing Alfv\'{e}n wave turbulence in the upper atmosphere
\citep{2011ApJ...736....3V}, the simple process outlined in this paper is suggested to be considered as a viable mechanism that can contribute to the heating of magnetic elements.
To be more precise, we conjecture that more than one mechanism of wave production is effective during the excursion of a magnetic element in the erratic intergranular lane, and result in temporal variations of the overall emission that occur at chromospheric heights over these magnetic elements.

\begin{acknowledgments}
\begin{small}This work was supported by the Science and Technology Facilities Council (STFC), UK. We thank Aad van Ballegooijen for providing the initial model described in the Appendix and for the suggestions and comments which significantly improved the paper. The authors are grateful to Oskar Steiner for carefully reading the manuscript and offering his valuable remarks. RE acknowledges M. Ker\'ay for patient encouragement and is also grateful
to NSF, Hungary (OTKA, Ref. No. 483133). We would like to thank the anonymous referee for his/her detailed comments, which helped us to improve the paper.\end{small}
\end{acknowledgments}

\appendix
\section{Initial Magneto-Hydrostatic Model}

In this section we develop a 3D model for the magnetic field ${\bf B}({\bf r})$ and plasma parameters of a photospheric magnetic flux tube.  A Cartesian reference frame $(x,y,z)$ is used with $z$ the height above the base of the photosphere (i.e., the level where the optical depth $\tau_{5000} = 1$). The computational domain is given by $x = [-L,L]$, $y = [-L,L]$, and $z=[0,L_z]$, where $L$ and $L_z$ are the half-width and height of the domain, respectively. The axis of the flux tube is vertical and lies along the $z$-axis of the reference frame. The flux tube is symmetric with respect to the planes $x = 0$ and $y = 0$, so for the purpose of computing the initial equilibrium field only one quarter of the domain needs to be considered: $x = [0,L]$ and $y = [0,L]$. At low heights the flux tube has a nearly circular cross-section with radius $r_0 < L$, and is surrounded by a nearly field-free medium. However, the gas pressure decreases with height $z$, causing the cross-section of the tube to increase with height, and at larger heights the radius of the flux tube becomes comparable to the half-width $L$ of the domain. To simulate the effect of the neighboring flux tubes, we assume that the normal component of ${\bf B}$ vanishes at the side walls of the domain, so that the field lines are forced up along the side walls. In this paper we use $L = 0.5$ Mm, corresponding to a total width of the flux tube of 1 Mm at large height.\\

The magnetic field ${\bf B} ({\bf r})$ is written in terms of Euler potentials:
\begin{equation}
{\bf B} = \bnabla u \times \bnabla v , \label{eq:euler}
\end{equation}
where $u ({\bf r})$ and $v ({\bf r})$ are constant along field lines
(${\bf B} \cdot \bnabla u = {\bf B} \cdot \bnabla v = 0$). At the base of the model the vertical component of the field has a Gaussian profile,
\begin{equation}
B_z (x,y,0) = B_0 \exp [-(x^2+y^2)/r_0^2 ] , \label{eq:Bz0}
\end{equation}
where $r_0$ is the $1/e$-width of the profile (we use $r_0 = 120$ km) and $B_0$ is the peak field strength. The Euler potentials at the base are chosen such that $u(x,y,0)$ depends only on $x$ and $v(x,y,0)$ depends only on $y$:
\begin{equation}
u(x,y,0) = G(x) ~~~ \hbox{and} ~~~ v(x,y,0) = G(y) ,
\end{equation}
where
\begin{equation}
G(x) \equiv \sqrt{B_0} \int_0^x \exp \left( -\frac{s^2} {r_0^2}
\right) ~ ds ,
\end{equation}
which is consistent with equations (\ref{eq:euler}) and (\ref{eq:Bz0}). The side boundary conditions are
\begin{equation}
u(0,y,z)=0, ~~~~ u(L,y,z)=G_{\rm max} , ~~~~ v(x,0,z)=0 , ~~~~
v(x,L,z)=G_{\rm max} ,
\end{equation}
where $G_{\rm max} = G(L)$. At the upper boundary no conditions on $u(x,y,L_z)$ and $v(x,y,L_z)$ are imposed, allowing the field lines to freely move on the upper boundary. Hence, there should be no build-up of field-aligned electric currents inside the domain. The interior of the domain is assumed to be in magnetostatic equilibrium: 
\begin{equation}
- \nabla p + \rho {\bf g} + \frac{1}{4\pi} ( \nabla \times {\bf B} )
\times {\bf B} = 0 , \label{eq:force}
\end{equation}
where ${\bf g} \equiv - g \hat{\bf z}$ is the gravitational acceleration, $p ({\bf r})$ is the plasma pressure, and $\rho ({\bf r})$ is the mass density. The third term on the left-hand side describes the Lorentz force due to distributed electric currents at the interface between the flux tube and its surroundings. The gas pressure and density are given by
\begin{eqnarray}
p(u,v,z) & = & p_{\rm int}(z) \left[ 1 + \beta_0^{-1} f(u,v) \right] ,
\label{eq:pres} \\
\rho (u,v,z) & = & \rho_{\rm int}(z) \left[ 1 + \beta_0^{-1} f(u,v)
\right] , \label{eq:dens}
\end{eqnarray}
where $p_{\rm int} (z)$ and $\rho_{\rm int}(z)$ are the internal pressure and density along the axis of the flux tube, $\beta_0$ is the ratio of gas- and magnetic pressures at the origin ($z=0$), and $f(u,v)$ varies smoothly from zero on the flux tube axis to $f(u,v)=1$ in the external medium and at the side walls. The internal density is given by $\rho_{\rm int} (z) = - g^{-1} dp_{\rm int}/dz$, so there is hydrostatic equilibrium along field lines. Note that $p/\rho$ depends only on height $z$, so the temperature is constant in horizontal planes. The internal pressure as function of height is approximated as a sum of two exponentials:
\begin{equation}
p_{\rm int} (z) = p_1 \exp(-z/H_1) + p_2 \exp(-z/H_2) ,
\end{equation}
with a photospheric pressure scale height $H_1$ = 110 km ($p_1 = 4.2 \times 10^4$ $\rm dyne ~ cm^{-2}$), and a chromospheric scale height $H_2$ = 220 km ($p_2 = 10^3$ $\rm dyne ~ cm^{-2}$). The maximum of the magnetic field strength at the foot point is estimated using the thin tube approximation: $B_0 = \sqrt{8 \pi p_{\rm int}(0) / \beta_0}$. To obtain kilogauss fields in the photosphere we use $\beta_0 = 0.5$, so that the external gas pressure is three times the internal gas pressure. The function $f(u,v)$ is given by
\begin{equation}
f(u,v) = F(A) \equiv (4A - A^4 ) /3 ,
\end{equation}
where
\begin{equation}
A(u,v) = 1 - \cos \left( \frac{\pi u}{2 G_{\rm max}} \right)
\cos \left( \frac{\pi v}{2 G_{\rm max}} \right) .
\end{equation}
Note that $A(u,v)$ varies from $A=0$ on the axis ($u=v=0$) to $A=1$ in the field-free medium and on the side boundaries ($u=G_{\rm max}$ or $v=G_{\rm max}$), and that $dF/dA = 0$ at $A=1$. Hence, there is a smooth transition of the gas pressure from the interior to the exterior of the flux tube, and the electric currents at the interface are finite.\\

Inserting expressions (\ref{eq:euler}), (\ref{eq:pres}) and (\ref{eq:dens}) into equation (\ref{eq:force}) yields two coupled equations for the Euler potentials $u$ and $v$:
\begin{eqnarray}
\frac{1}{4 \pi} ( \bnabla \times {\bf B} ) \cdot \bnabla v
- p_{\rm int}(z) \beta_0^{-1} \frac{\partial f} {\partial u} & = & 0 , \\
\frac{1}{4 \pi} ( \bnabla \times {\bf B} ) \cdot \bnabla u
+ p_{\rm int}(z) \beta_0^{-1} \frac{\partial f} {\partial v} & = & 0 .
\end{eqnarray}
The solution of these equations can be obtained by solving a variational problem for the Lagrangian,
\begin{equation}
W \equiv \int_0^L \int_0^L \int_0^{L_z} \left[
\frac{| \bnabla u \times \bnabla v |^2} {8 \pi} - p_{\rm int}(z)
\beta_0^{-1} f(u,v) \right] ~ dx ~ dy ~ dz , \label{eq:func}
\end{equation}
where the variations in $(u,v)$ are subject to the above-mentioned boundary conditions. To solve this variational problem, we set up a grid of $50 \times 50 \times 100$ cubic cells covering the interior of the computational domain, and ghost cells are added at the side boundaries. The Euler potentials $u_{ijk}$ and $v_{ijk}$ are defined at the cell corners ($i$, $j$ and $k$ are indices on the grid), and the derivatives in equation (\ref{eq:func}) are evaluated using second-order finite differences. The Lagrangian $W(u_{ijk},v_{ijk})$ is then a function of the potentials at all interior grid points. We minimize this quantity using the conjugent-gradient method
\citep{1992nrfa.book.....P}. 
Once the solution is obtained, the magnetic field, pressure and density are evaluated at cell centers, and the computed values are mirrored to obtain a 3D magnetostatic model of the entire flux tube. This model is used as the initial condition for the 3D MHD calculations.

\bibliographystyle{apj}
\bibliography{ms}

\begin{thebibliography}{42}
\expandafter\ifx\csname natexlab\endcsname\relax\def\natexlab#1{#1}\fi

\bibitem[{{Berger} \& {Title}(1996)}]{1996ApJ...463..365B}
{Berger}, T.~E. \& {Title}, A.~M. 1996, \apj, 463, 365

\bibitem[{{Bonet} {et~al.}(2008){Bonet}, {M{\'a}rquez}, {S{\'a}nchez Almeida},
  {Cabello}, \& {Domingo}}]{2008ApJ...687L.131B}
{Bonet}, J.~A., {M{\'a}rquez}, I., {S{\'a}nchez Almeida}, J., {Cabello}, I., \&
  {Domingo}, V. 2008, \apjl, 687, L131

\bibitem[{{Bonet} {et~al.}(2010){Bonet}, {M{\'a}rquez}, {S{\'a}nchez Almeida},
  {Palacios}, {Mart{\'{\i}}nez Pillet}, {Solanki}, {del Toro Iniesta},
  {Domingo}, {Berkefeld}, {Schmidt}, {Gandorfer}, {Barthol}, \&
  {Kn{\"o}lker}}]{2010ApJ...723L.139B}
{Bonet}, J.~A., {M{\'a}rquez}, I., {S{\'a}nchez Almeida}, J., {Palacios}, J.,
  {Mart{\'{\i}}nez Pillet}, V., {Solanki}, S.~K., {del Toro Iniesta}, J.~C.,
  {Domingo}, V., {Berkefeld}, T., {Schmidt}, W., {Gandorfer}, A., {Barthol},
  P., \& {Kn{\"o}lker}, M. 2010, \apjl, 723, L139

\bibitem[{{Bray} \& {Loughhead}(1974)}]{1974soch.book.....B}
{Bray}, R.~J. \& {Loughhead}, R.~E. 1974, {The solar chromosphere}, ed. {Bray,
  R.~J.~\& Loughhead, R.~E.}

\bibitem[{{Cranmer} \& {van Ballegooijen}(2005)}]{2005ApJS..156..265C}
{Cranmer}, S.~R. \& {van Ballegooijen}, A.~A. 2005, \apjs, 156, 265

\bibitem[{{Erd{\'e}lyi} {et~al.}(2007){Erd{\'e}lyi}, {Malins}, {T{\'o}th}, \&
  {de Pontieu}}]{2007A&A...467.1299E}
{Erd{\'e}lyi}, R., {Malins}, C., {T{\'o}th}, G., \& {de Pontieu}, B. 2007,
  \aap, 467, 1299

\bibitem[{{Fedun} {et~al.}(2009){Fedun}, {Erd{\'e}lyi}, \&
  {Shelyag}}]{2009SoPh..258..219F}
{Fedun}, V., {Erd{\'e}lyi}, R., \& {Shelyag}, S. 2009, \solphys, 258, 219

\bibitem[{{Fedun} {et~al.}(2011{\natexlab{a}}){Fedun}, {Shelyag}, \&
  {Erd{\'e}lyi}}]{2011ApJ...727...17F}
{Fedun}, V., {Shelyag}, S., \& {Erd{\'e}lyi}, R. 2011{\natexlab{a}}, \apj, 727,
  17

\bibitem[{{Fedun} {et~al.}(2011{\natexlab{b}}){Fedun}, {Shelyag}, {Verth},
  {Mathioudakis}, \& {Erd{\'e}lyi}}]{2011AnGeo..29.1029F}
{Fedun}, V., {Shelyag}, S., {Verth}, G., {Mathioudakis}, M., \& {Erd{\'e}lyi},
  R. 2011{\natexlab{b}}, Annales Geophysicae, 29, 1029

\bibitem[{{Fedun} {et~al.}(2011{\natexlab{c}}){Fedun}, {Verth}, {Jess}, \&
  {Erd{\'e}lyi}}]{2011ApJ...740L..46F}
{Fedun}, V., {Verth}, G., {Jess}, D.~B., \& {Erd{\'e}lyi}, R.
  2011{\natexlab{c}}, \apjl, 740, L46+

\bibitem[{{Hasan} \& {Kalkofen}(1999)}]{1999ApJ...519..899H}
{Hasan}, S.~S. \& {Kalkofen}, W. 1999, \apj, 519, 899

\bibitem[{{Hasan} {et~al.}(2000){Hasan}, {Kalkofen}, \& {van
  Ballegooijen}}]{2000ApJ...535L..67H}
{Hasan}, S.~S., {Kalkofen}, W., \& {van Ballegooijen}, A.~A. 2000, \apjl, 535,
  L67

\bibitem[{{Hasan} \& {van Ballegooijen}(2008)}]{2008ApJ...680.1542H}
{Hasan}, S.~S. \& {van Ballegooijen}, A.~A. 2008, \apj, 680, 1542

\bibitem[{{Hasan} {et~al.}(2005){Hasan}, {van Ballegooijen}, {Kalkofen}, \&
  {Steiner}}]{2005ApJ...631.1270H}
{Hasan}, S.~S., {van Ballegooijen}, A.~A., {Kalkofen}, W., \& {Steiner}, O.
  2005, \apj, 631, 1270

\bibitem[{{Jess} {et~al.}(2009){Jess}, {Mathioudakis}, {Erd{\'e}lyi},
  {Crockett}, {Keenan}, \& {Christian}}]{2009Sci...323.1582J}
{Jess}, D.~B., {Mathioudakis}, M., {Erd{\'e}lyi}, R., {Crockett}, P.~J.,
  {Keenan}, F.~P., \& {Christian}, D.~J. 2009, Science, 323, 1582

\bibitem[{{Kato} {et~al.}(2011){Kato}, {Steiner}, {Steffen}, \&
  {Suematsu}}]{2011ApJ...730L..24K}
{Kato}, Y., {Steiner}, O., {Steffen}, M., \& {Suematsu}, Y. 2011, \apjl, 730,
  L24+

\bibitem[{{Keys} {et~al.}(2011){Keys}, {Mathioudakis}, {Jess}, {Shelyag},
  {Crockett}, {Christian}, \& {Keenan}}]{2011ApJ...740L..40K}
{Keys}, P.~H., {Mathioudakis}, M., {Jess}, D.~B., {Shelyag}, S., {Crockett},
  P.~J., {Christian}, D.~J., \& {Keenan}, F.~P. 2011, \apjl, 740, L40

\bibitem[{{Khomenko} \& {Cally}(2012)}]{2012ApJ...746...68K}
{Khomenko}, E. \& {Cally}, P.~S. 2012, \apj, 746, 68

\bibitem[{{Khomenko} {et~al.}(2008){Khomenko}, {Collados}, \&
  {Felipe}}]{2008SoPh..251..589K}
{Khomenko}, E., {Collados}, M., \& {Felipe}, T. 2008, \solphys, 251, 589

\bibitem[{{Kitiashvili} {et~al.}(2011){Kitiashvili}, {Kosovichev}, {Mansour},
  \& {Wray}}]{2011ApJ...727L..50K}
{Kitiashvili}, I.~N., {Kosovichev}, A.~G., {Mansour}, N.~N., \& {Wray}, A.~A.
  2011, \apjl, 727, L50+

\bibitem[{{Malins} \& {Erd{\'e}lyi}(2007)}]{2007SoPh..246...41M}
{Malins}, C. \& {Erd{\'e}lyi}, R. 2007, \solphys, 246, 41

\bibitem[{{Moll} {et~al.}(2011){Moll}, {Cameron}, \&
  {Sch{\"u}ssler}}]{2011A&A...533A.126M}
{Moll}, R., {Cameron}, R.~H., \& {Sch{\"u}ssler}, M. 2011, \aap, 533, A126

\bibitem[{{Musielak} {et~al.}(1995){Musielak}, {Rosner}, {Gail}, \&
  {Ulmschneider}}]{1995ApJ...448..865M}
{Musielak}, Z.~E., {Rosner}, R., {Gail}, H.~P., \& {Ulmschneider}, P. 1995,
  \apj, 448, 865

\bibitem[{{Musielak} {et~al.}(1989){Musielak}, {Rosner}, \&
  {Ulmschneider}}]{1989ApJ...337..470M}
{Musielak}, Z.~E., {Rosner}, R., \& {Ulmschneider}, P. 1989, \apj, 337, 470

\bibitem[{{Musielak} {et~al.}(2000){Musielak}, {Rosner}, \&
  {Ulmschneider}}]{2000ApJ...541..410M}
---. 2000, \apj, 541, 410

\bibitem[{{Musielak} {et~al.}(2002){Musielak}, {Rosner}, \&
  {Ulmschneider}}]{2002ApJ...573..418M}
---. 2002, \apj, 573, 418

\bibitem[{{Press} {et~al.}(1992){Press}, {Teukolsky}, {Vetterling}, \&
  {Flannery}}]{1992nrfa.book.....P}
{Press}, W.~H., {Teukolsky}, S.~A., {Vetterling}, W.~T., \& {Flannery}, B.~P.
  1992, {Numerical recipes in FORTRAN. The art of scientific computing}

\bibitem[{{Shelyag} {et~al.}(2008){Shelyag}, {Fedun}, \&
  {Erd{\'e}lyi}}]{2008A&A...486..655S}
{Shelyag}, S., {Fedun}, V., \& {Erd{\'e}lyi}, R. 2008, \aap, 486, 655

\bibitem[{{Shelyag} {et~al.}(2011{\natexlab{a}}){Shelyag}, {Fedun}, {Keenan},
  {Erd{\'e}lyi}, \& {Mathioudakis}}]{2011AnGeo..29..883S}
{Shelyag}, S., {Fedun}, V., {Keenan}, F.~P., {Erd{\'e}lyi}, R., \&
  {Mathioudakis}, M. 2011{\natexlab{a}}, Annales Geophysicae, 29, 883

\bibitem[{{Shelyag} {et~al.}(2011{\natexlab{b}}){Shelyag}, {Keys},
  {Mathioudakis}, \& {Keenan}}]{2011A&A...526A...5S}
{Shelyag}, S., {Keys}, P., {Mathioudakis}, M., \& {Keenan}, F.~P.
  2011{\natexlab{b}}, \aap, 526, A5+

\bibitem[{{Steiner} {et~al.}(2010){Steiner}, {Franz}, {Bello Gonz{\'a}lez},
  {Nutto}, {Rezaei}, {Mart{\'{\i}}nez Pillet}, {Bonet Navarro}, {del Toro
  Iniesta}, {Domingo}, {Solanki}, {Kn{\"o}lker}, {Schmidt}, {Barthol}, \&
  {Gandorfer}}]{2010ApJ...723L.180S}
{Steiner}, O., {Franz}, M., {Bello Gonz{\'a}lez}, N., {Nutto}, C., {Rezaei},
  R., {Mart{\'{\i}}nez Pillet}, V., {Bonet Navarro}, J.~A., {del Toro Iniesta},
  J.~C., {Domingo}, V., {Solanki}, S.~K., {Kn{\"o}lker}, M., {Schmidt}, W.,
  {Barthol}, P., \& {Gandorfer}, A. 2010, \apjl, 723, L180

\bibitem[{{Taroyan} \& {Erd{\'e}lyi}(2009)}]{2009SSRv..149..229T}
{Taroyan}, Y. \& {Erd{\'e}lyi}, R. 2009, \ssr, 149, 229

\bibitem[{{T{\'o}th}(1996)}]{1996ApL&C..34..245T}
{T{\'o}th}, G. 1996, Astrophysical Letters Communications, 34, 245

\bibitem[{{Utz} {et~al.}(2010){Utz}, {Hanslmeier}, {Muller}, {Veronig},
  {Ryb{\'a}k}, \& {Muthsam}}]{2010A&A...511A..39U}
{Utz}, D., {Hanslmeier}, A., {Muller}, R., {Veronig}, A., {Ryb{\'a}k}, J., \&
  {Muthsam}, H. 2010, \aap, 511, A39

\bibitem[{{van Ballegooijen} {et~al.}(2011){van Ballegooijen}, {Asgari-Targhi},
  {Cranmer}, \& {DeLuca}}]{2011ApJ...736....3V}
{van Ballegooijen}, A.~A., {Asgari-Targhi}, M., {Cranmer}, S.~R., \& {DeLuca},
  E.~E. 2011, \apj, 736, 3

\bibitem[{{Verth} {et~al.}(2010){Verth}, {Erd{\'e}lyi}, \&
  {Goossens}}]{2010ApJ...714.1637V}
{Verth}, G., {Erd{\'e}lyi}, R., \& {Goossens}, M. 2010, \apj, 714, 1637

\bibitem[{{Vigeesh} {et~al.}(2009){Vigeesh}, {Hasan}, \&
  {Steiner}}]{2009A&A...508..951V}
{Vigeesh}, G., {Hasan}, S.~S., \& {Steiner}, O. 2009, \aap, 508, 951

\bibitem[{{Vigeesh} {et~al.}(2011){Vigeesh}, {Steiner}, \&
  {Hasan}}]{2011SoPh..tmp..349V}
{Vigeesh}, G., {Steiner}, O., \& {Hasan}, S.~S. 2011, \solphys, 273, 15

\bibitem[{{Wedemeyer-B{\"o}hm} \& {Rouppe van der
  Voort}(2009)}]{2009A&A...507L...9W}
{Wedemeyer-B{\"o}hm}, S. \& {Rouppe van der Voort}, L. 2009, \aap, 507, L9

\bibitem[{{Wedemeyer-B{\"o}hm} {et~al.}(2012){Wedemeyer-B{\"o}hm}, {Scullion},
  {Steiner}, {Rouppe van der Voort}, {de la Cruz Rodriguez}, {Fedun}, \&
  {Erd{\'e}lyi}}]{2012Nature_Wedemeyer}
{Wedemeyer-B{\"o}hm}, S., {Scullion}, E., {Steiner}, O., {Rouppe van der
  Voort}, L., {de la Cruz Rodriguez}, J., {Fedun}, V., \& {Erd{\'e}lyi}, R.
  2012, Nature, 486, 505

\end{thebibliography}

\end{document}